\documentclass[aps,physrev,twocolumns,groupedaddress,amsmath,amssymb,10pt]{revtex4-1}

\usepackage{graphicx}
\usepackage{dcolumn}
\usepackage{bm}
\usepackage{lipsum}
\usepackage{color}
\usepackage{array}
\usepackage{relsize}
\usepackage{multirow}
\usepackage{booktabs}
\usepackage{tcolorbox}
\usepackage{subcaption}

\usepackage{hyperref}

\begin{document}


\title{Soft and semihard components of multiplicity distributions in the $k_T$ 
factorization approach}

\author{H. R. Martins-Fontes}
 \email{henriquemartinsfontes@usp.br}
\author{F. S. Navarra}%
 \email{navarra@if.usp.br}
\affiliation{Instituto de Física, Universidade de São Paulo, Rua do Matão 1371, 
Cidade Universitária, - 05508-090,
São Paulo, SP, Brazil.}

\date{\today}

\begin{abstract}
Particle production in hadronic collisions can be studied in the low-momentum (soft) and high-momentum (hard) transfer regimes. While the latter can be well understood within perturbative QCD the former contains non-perturbative effects which cannot
be calculated from first principles. There is also an intermediate regime called semihard, in which the momentum transfer runs 
typically from $~1$ to $~10$ GeV. As the hadron-hadron  collision energy increases, we expect to see a relative growth of the 
number of semihard events. It has been conjectured that this growth would be the cause of some changes observed in the multiplicity 
distributions measured in proton - proton collisions. In this note we revisit the separation between soft and semihard events using 
the formalism of $k_T$ factorization. The separation is implemented through the introduction of a scale that is the cutoff $\Lambda$ in the transverse momentum of the produced gluon and allows us to compute the average number of particles produced in each regime.  These 
numbers are used as input in the double negative binomial fit of data, from which we can extract correlations between the fraction of 
semihard events and the violation of KNO scaling. 
\end{abstract}

\maketitle

\section{Introduction}

The multiplicity distribution (MD) of charged particles measured in high energy proton-proton collisions is one the observables that is not yet well understood in QCD. For a comprehensive review, see \cite{Grosse-Oetringhaus:2009eis}.  For the multiplicities
of particles in jets, in the limit
of very high energies and very high multiplicities, there are robust predictions of perturbative QCD (pQCD) \cite{Dokshitzer:2025owq}. 
For other types of events, there are significant non-perturbative effects and we cannot perform calculations from first principles. 
In the absence of a straightforward connection between theory and data, we have to resort 
to models.  However even this task is complex because of the non-perturbative nature of hadronization. The "top-down" models, 
such as those based on the Color Glass Condensate (CGC)  \cite{Gelis:2010nm} approach, or event generators, such as PYTHIA
\cite{Sjostrand:2006za,Bierlich:2020naj}, give only an approximate description of the measured MD's which exhibit a complex  structure.  Many works have taken a "bottom-up" approach, i.e. trying first to find empirical regularities in the data, obtaining successful parametrizations of these data and then searching for scaling laws and their violation. After this first step, one tries to connect the 
empirical observations with the theory. Along this line the fits with the Negative Binomial Distribution (NBD) were very successful for 
many years and specially for lower energies data.  But, at  the $\sqrt{s} = 900$ GeV SPS energy, the NBD started to fail in describing the measured multiplicity distributions, which exhibited a shoulder-shaped structure \cite{Grosse-Oetringhaus:2009eis}.  

For LHC energies, a single NBD fit turned out to be problematic.    
In the early ALICE data \cite{ALICE:2010cin}, for the class of non-single diffractive (NSD) events, no strong deviation from a single 
NBD fit was observed
at $\sqrt{s} = 0.9$ and $2.36$ TeV for $|\eta| < 1$, but a hint of a substructure appeared at $|\eta| < 1.3$. 
However, in \cite{ALICE:2015olq} for all event classes, already at $\sqrt{s} = 0.9$ TeV, the single NBD fits started to diverge
from the data at the highest multiplicities for $|\eta| < 0.5$ and $1.0$.  Even more significant departures from
the single NBD started at  $\sqrt{s} = 2.76$ TeV.

A fruitful idea was to interpret the shoulder-like  structure observed in the data
as coming from a superposition of two negative binomial distributions (called DNBD). The two NBDs could be associated to two different sources or two different types of processes. In the first case, as advocated, for example, in 
\cite{Fowler:1988jz,Fowler:1986nv}, the sources are well separated in rapidity. In the second case, the two NBDs would come from soft and semihard processes. This latter idea was advanced long ago in \cite{Giovannini:1998zb},  further developed 
in  \cite{Ghosh:2012xh} and more recently in \cite{Zborovsky:2018vyh}. A review on this subject can be found in \cite{Giovannini:2004yk}.
For the TeV region, the expectation was that at increasing energies the semihard component would become increasingly more important than the soft one.  In parallel, a violation of the Koba-Nielsen-Olsen (KNO) scaling  \cite{Koba:1972ng} 
was observed, suggesting a connection between the two findings. The latest LHC data  \cite{acharya2023multiplicity, ALICE:2022kol}  
confirmed  that for narrow pseudorapidity windows a single NBD was enough to reproduce the observed distributions and
also that KNO scaling was valid.  For larger windows, two NBDs were needed and KNO scaling was violated.

In \cite{Ghosh:2012xh,Giovannini:2004yk} the difference between soft and semihard processes was only that the latter yield larger mean multiplicities.  In the present work, using the $k_T$ factorization formalism  \cite{Gribov:1983fc,Kovchegov:2012mbw},  we can make a better separation with the proper introduction of an energy scale. Moreover, the introduction of this formalism reduces the number of parameters.  As will be seen, 
we will obtain fits with very good quality and we will extract from them, with less ambiguity,  the behavior of the parameter $\alpha$ with the 
energy $\sqrt{s}$ and with the size of the pseudorapidity window.  This parameter represents the fraction of soft events. 

\section{The $k_T$ factorization approach}

At high energy collisions between hadrons or heavy ions, a large number of gluons is produced due to the dense gluon population in the hadron wave function. Using the CGC formalism, particle production in high-energy hadronic collisions can be studied through the $k_T$ factorization approach, where the inclusive production cross-section is given by \cite{Kovchegov:2012mbw,Wang:2013gzd}:
\begin{equation}
\label{Kt-energia}
    E \frac{d \sigma} {d^3 p}= K \frac{4 \pi N_{c}} {N_{c}^{2}-1} \alpha_{s}(Q^2)\; \frac{1} {p_{{\perp}}^{2}} \;  
    \int^{p_{{\perp}}} \, d k_{{\perp}}^{2} \; \varphi_{1} ( x_{1}, k_{{\perp}}^{2} ) \; \varphi_{2} ( x_{2}, ( p-k )_{{\perp}}^{2} ),
\end{equation}
where $x_{1,2} = \frac{p_{\perp}}{\sqrt{s}}\exp{(\pm y)}$, $\sqrt{s}$ is the center of mass energy, $N_c$ is the number of colors and $K$ is a normalization factor that describes the conversion of partons to hadrons. $\varphi$ is the  unintegrated gluon distribution (UGD) of a proton and it can be related to the gluon density by
\begin{equation}
\label{int-gluon}
    xG(x,\mu^2) = \int^{\mu^2} d k_{{\perp}}^{2} \varphi ( x, k_{{\perp}}^{2} ).
\end{equation}
The multiplicity per unit of rapidity can be computed by integrating (\ref{Kt-energia}) over $p_{\perp}$
\begin{equation}
    \frac{dN}{dy} = \frac{1}{S} \int_{p_{\perp{min}}}^{p_{\perp{max}}} d^2 p_{\perp}  E \frac{d \sigma} {d^3 p},
    \label{sep} 
\end{equation} 
where $S$ is a typical interaction area and the minimum and maximum value of $p_{\perp}$ are determined by the experimental conditions. In (\ref{sep}), the  main contribution is given by two regions of integration over $k_{\perp}$: $k_{\perp} \ll p_{\perp}$ and $\lvert 
p_{\perp} - k_{\perp} \rvert \ll p_{\perp}$. Hence we can rewrite it as 
\cite{kharzeev2005color}:

\begin{align}
    \frac{dN}{dy} &=  \frac{K}{S} \frac{4 \pi N_{c}} {N_{c}^{2}-1} \; \int_{p_{\perp{min}}}^{p_{\perp{max}}} \frac{d p_{\perp}^2} {p_{{\perp}}^{2}} \alpha_s(Q^2) \; \left[ \varphi_1(x_1,p_{\perp})\int^{p_{\perp}} dk_{\perp}^2 \varphi_2(x_2,k_{\perp}) + \varphi_2(x_2,p_{\perp})\int^{p_{\perp}} dk_{\perp}^2 \varphi_1(x_1,k_{\perp})\right] \nonumber \\
    &= \frac{K}{S} \frac{4 \pi N_{c}} {N_{c}^{2}-1} \; \int_{p_{\perp{min}}}^{p_{\perp{max}}} \frac{d p_{\perp}^2} {p_{{\perp}}^{4}} \alpha_s(Q^2) xG_2(x_2, p_{\perp}^2)xG_1(x_1, p_{\perp}^2),
    \label{kt-final} 
\end{align} 
where the equation in the first line was integrated by parts and use was made of  (\ref{int-gluon}). The multiplicative constants $K$ and $S$ can be grouped in a single parameter which is fixed by fitting the available pseudo-rapidity distributions. For the UGD, we adopt the one from the GBW model 
\cite{Golec-Biernat:1998zce,Golec-Biernat:2017lfv}  because it comes from  a simple and well-established dipole parameterization, 
which it is known for its success in the description of a  wealth of experimental data on particle production, including both hard and soft processes.  For the purposes of our work the GBW model is specially appropriate because it reproduces very 
well the DIS data on $F_2$ at very low scales, which are consistent with the low transverse momenta of the bulk of produced particles at the LHC. 
At these scales the UGD derived from the GBW model is also consistent with much more sophisticated UGDs  \cite{Moriggi:2020zbv}. 
For a recent account of the successes of the GBW model, see \cite{Henkels:2023plt} and references therein. The GBW UGD reads: 

\begin{equation}
\label{GBW-UGD}
    \varphi_{1,2}(x_{1,2};k_{\perp}^2) = \frac{3\sigma_0}{4\pi^2\alpha_s(Q_{s_{1,2}}^2)} \frac{k_{\perp}^2}{Q_{s_{1,2}}^2} \exp{\left(-\frac{k_{\perp}^2}{Q_{s_{1,2}}^2}\right)},
\end{equation}
where the saturation scale $Q_{s}^2$ is defined as
\begin{equation}
    Q_{s_{1,2}}^2 = Q_0^2\left( x_0 \frac{\sqrt{s}}{Q_0}e^{\pm y} \right)^{\bar{\lambda}}.
\end{equation}
The parameters were fixed as in \cite{Wang:2013gzd}: $Q_0 = 0.6 \, \text{GeV}$, $x_0 = 0.01$, $\bar{\lambda} = 0.205$, and $\sigma_0 = 23 \, \text{mb}$. The running coupling constant $\alpha_s$ follows the one-loop approximation and is assumed to freeze at 0.52:
\begin{equation}
    \alpha_s(Q^2) = \min\left[ \frac{12\pi}{27\ln{\left( 
  \frac{Q^2}{\Lambda_{QCD}^2} \right)}}, 0.52 \right],
\end{equation}
with $\Lambda_{QCD} = 0.226 \, \text{GeV}$. The scale \(Q^2\) at which the strong coupling \(\alpha_s(Q^2)\) is evaluated in Eq. (\ref{kt-final}) is defined as
\begin{equation}
    Q^2 = \max{[p_{\perp}^2,\min{(Q^2_{s,1},Q^2_{s,2} )}]}.
\end{equation}
The UGD (\ref{GBW-UGD}) leads to the gluon distribution function expressed as
\begin{equation}
    xG_{1,2}(x_{1,2};p_{\perp}^2) =  \frac{3\sigma_0}{4\pi^2\alpha_s(Q_{s_{1,2}}^2)} \left(Q_{s_{1,2}}^2 - \exp{\left(-\frac{p_{\perp}^2}{Q_{s_{1,2}}^2}\right)}(p_{\perp}^2 + Q_{s_{1,2}}^2)  \right)(1 - x_{1,2})^4.
\end{equation}
Here, as in \cite{kharzeev2005color}, the factor $(1 - x)^4$ is introduced to account for large-$x$ effects.
In expression  Eq. (\ref{Kt-energia}),   and also in  Eq. (\ref{kt-final}), one gluon with $p_{\perp}$ is produced from the fusion of a gluon with $k_{\perp}$ and another gluon with
$p_{\perp} - k_{\perp}$. Therefore the momentum $p_{\perp}$ is, to a good approximation, the minimum value of the momentum flowing in the three gluon vertex, i.e. the one to be used in the running coupling constant. For $p_{\perp}$ larger than a few GeV, the coupling $\alpha_s(Q^2) $ 
becomes sufficiently smaller than one and we are in the perturbative domain. We will thus follow the literature and call these events 
"semihard". The events with lower values of $p_{\perp}$  will be called "soft". Accordingly the total multiplicity per unit of rapidity is 
the sum of the contributions from  the soft and semihard events. The separation is achieved by introducing a cutoff $\Lambda$ in the integral over $p_{\perp}$, which defines the two contributions:
\begin{eqnarray}
\label{sep_final}
    \frac{dN}{dy} &=& \frac{K}{S} \frac{4 \pi N_{c}} {N_{c}^{2}-1} \left[\; \int_{p_{\perp{min}}}^{\Lambda} \frac{d p_{\perp}^2} {p_{{\perp}}^{4}} \alpha_s(Q^2) xG_2(x_2, p_{\perp}^2)xG_1(x_1, p_{\perp}^2) + 
     \int_{{\Lambda}}^{p_{\perp{max}}} \frac{d p_{\perp}^2} {p_{{\perp}}^{4}} \alpha_s(Q^2) xG_2(x_2, p_{\perp}^2)xG_1(x_1, p_{\perp}^2)\right]   \nonumber \\
                 &=& \frac{dN_s}{dy} + \frac{dN_{sh}}{dy}.
\end{eqnarray}
where $ N_s $ and $N_{sh}$ are the soft and semihard components respectively. 
In the context of the CGC formalism, a similar separation criterion was introduced 
in \cite{Carvalho:2007cf}. 
To calculate the pseudo-rapidity distributions, one should rewrite expression  (\ref{sep_final}) using the transformation 
\begin{equation}
    y ( \eta)=\frac{1} {2} \operatorname{l o g} \frac{\sqrt{\operatorname{c o s h}^{2} \eta+\mu^{2}}+\operatorname{s i n h} \eta} {\sqrt{\operatorname{c o s h}^{2} \eta+\mu^{2}}-\operatorname{s i n h} \eta},
\end{equation}
where the scale $\mu$ is given as \cite{Dumitru:2011wq}
\begin{equation}
    \mu( \sqrt{s} )=\frac{0. 2 4} {0. 1 3+0. 3 2 \ (\sqrt{s})^{0. 1 1 5}}.
\end{equation}
and the Jacobian reads:
\begin{equation}
J ( \eta)=\frac{\partial y} {\partial\eta}=\frac{\operatorname{c o s h} \eta} {\sqrt{\operatorname{c o s h}^{2} \eta+\mu^{2}}},
\end{equation}
Changing variables and integrating (\ref{sep_final}) over the pseudo-rapidity  we obtain the average multiplicity:
\begin{equation}
  N  \, = \,   N_s  \, + \,  N_{sh}.
\label{n-cgc}
\end{equation} 

We emphasize that the only free parameter is the constant $K/S$ which is going to be determined by using 
Eq. (\ref{sep_final}) to fit the measured pseudo-rapidity distributions.  It is also important to mention that 
Eq. (\ref{sep_final}) describes the gluon production and, after integration, the number of gluons produced. As done in 
previous works, we shall assume here that the number of produced partons is equal to the number of hadrons resulting 
from the hadronization of these partons ("parton-hadron duality"). The approximate equality between these numbers has
been (successfully) tested recently in \cite{Duan:2025ngi}, where the authors used PYTHIA to calculate the hadron multiplicities 
within jets with and without using the LUND prescription for hadronization. They found a small difference between the number of 
partons and the number of hadrons.

In Fig. \ref{fig1} we show the rapidity distributions measured by different 
collaborations of the LHC at different energies. The $k_T$ factorization distribution seems to be consistent with the data, and thus we believe that for the purposes of this work, using Eq. (\ref{sep_final}) is sufficient.

\section{Double Negative Binomial Distribution}

In \cite{Giovannini:1997ce}  Giovannini e Ugoccioni  proposed a two-component model combining two NBDs associated with two event classes: "soft" and "semihard" processes.  Recent analyses performed by the ALICE Collaboration showed that the double NBD fit yields a good description of the data on multiplicity distributions from 0.9 to 8 TeV \cite{ALICE:2015olq,acharya2017charged,acharya2023multiplicity} within several pseudo-rapidity windows.  In this work, we will use their model.  The DNBD is expressed as:
\begin{equation} 
\label{DNBD}
    P(n) = \lambda \; [ \; \alpha \; P(n,\langle n\rangle_s,k_s) + (1-\alpha) \; P(n,\langle n \rangle_{sh},k_{sh}) \; ],
\end{equation}
where
\begin{equation}
P(n,\langle n \rangle,k) = \frac{\Gamma(k+n)}{\Gamma(k) \, \Gamma(n+1)} \frac{\langle n \rangle^n  \, k^k}{(\langle n \rangle + k)^{n+k}}
\label{NB}.
\end{equation}
and where, in each NBD,  $\langle n \rangle$ is the average multiplicity and the parameter $k$ is related to the dispersion $D$ ($D^2 =  \langle n^2 \rangle
 - \langle n \rangle^2 $) through:
 \begin{equation}
\label{dispersion}
      \frac{D_i^2}{\langle n \rangle_i^2} = \frac{1}{ \langle n \rangle_i} + \frac{1}{k_i},
 \end{equation}
where $i=\{s,sh\}$. The parameter $\alpha$  is the fraction of soft events and consequently $(1-\alpha)$ is the fraction of semihard events;  $\langle n \rangle_s$ and $\langle n \rangle_{sh}$ 
are the mean multiplicities of soft and semihard events respectively; $k_s$ and $k_{sh}$ are the negative binomial parameters of the
soft and semihard components.  
The first few bins can not be reproduced by any NBD and they are related to a different production mechanism, not included in our formalism. Therefore, these bins were removed from the fitting process. To account for this, a normalization factor $\lambda$ was introduced \cite{ALICE:2015olq,acharya2017charged}. 
The average multiplicity obtained from  (\ref{DNBD}) can be expressed as:
\begin{equation}
    \langle n \rangle = \lambda\;[\;\alpha \;\langle n \rangle_s + (1-\alpha) \;\langle n \rangle_{sh}\;].
 \label{2nbd}
 \end{equation}
Comparing expression (\ref{2nbd}) with (\ref{n-cgc})  and recalling that $N = \langle n \rangle $  we conclude that:
 \begin{eqnarray}
\lambda \, \alpha  \, \langle n \rangle_s     &=& N_s  \nonumber \\  
\lambda \, (1-\alpha) \langle n \rangle_{sh}  &=& N_{sh}  
\label{sis}
 \end{eqnarray}
Since $N_s$ and $N_{sh}$ are known, these two equations reduce the number of free parameters from six to four: $\alpha$, $\lambda$, $k_s$,
and $k_{sh}$.  

\section{KNO scaling}

The Koba-Nielsen-Olesen (KNO) scaling \cite{Koba:1972ng} is a feature of MDs that asserts that, in the limit of $\langle n \rangle \to \infty$, the probability distributions $P(n)$ at different energies, once rescaled by $\langle n \rangle$, become copies of each other. In other words, if one plots $\langle n \rangle P(n)$ versus $n/\langle n \rangle$ for different energies, all curves collapse onto a universal function $\psi\left(n/\langle n \rangle\right)$:

\begin{equation}
    \langle n \rangle P(n)= \psi\left(\frac{n}{\langle n \rangle}\right).
\end{equation}
A direct consequence of the KNO scaling is that the total scaled variance, in the scaling limit, increases linearly with $\langle n \rangle$ 
\cite{Begun:2008fm,Mackowiak-Pawlowska:2014ipa,Chandra:2019imn}
\begin{equation}
\omega = \frac{D^2}{\langle n \rangle} = \kappa \,  \langle n \rangle,
\label{scav}
\end{equation}
with $\kappa = \, const > 0$.
In the case of the DNBD, using (\ref{2nbd}) we have:
\begin{eqnarray}
    D^2 &=&\lambda\left[ \alpha \langle n^2 \rangle_s + (1-\alpha)\langle n^2 \rangle_{sh} \right] - \langle n \rangle^2 
     = \lambda\left[ \alpha (D_s^2 +\langle n \rangle_s^2) + (1-\alpha)(D_{sh}^2 +\langle n \rangle_{sh}^2) \right] - \langle n \rangle^2 \\[1.5ex] \nonumber
        &=&\lambda\left[ \alpha \left(\langle n \rangle_s + \frac{\langle n \rangle_s^2}{k_s} + \langle n \rangle_s^2\right) + (1-\alpha)\left(\langle n \rangle_{sh} + \frac{\langle n \rangle_{sh}^2}{k_{sh}} + \langle n \rangle_{sh}^2 \right) \right] - \langle n \rangle^2 \\[1.5ex] \nonumber   
        &=&\langle n \rangle - \langle n \rangle^2 + \lambda\left[ \alpha \langle n \rangle_s^2\left( 1+\frac{1}{k_s} \right) + (1-\alpha)\langle n \rangle_{sh}^2\left( 1+\frac{1}{k_{sh}} \right) \right] 
\end{eqnarray}
where in the second line line we made use of (\ref{dispersion}). Dividing the above equation by $\langle n \rangle$ we obtain the scaled 
variance:
\begin{eqnarray}
  \omega =   \frac{D^2}{\langle n \rangle} &=& 1 - \langle n \rangle + {
  \frac{\langle n \rangle }{\langle n \rangle^2}\lambda\left[ \alpha \langle n \rangle_s^2\left( 1+\frac{1}{k_s} \right) + (1-\alpha)\langle n \rangle_{sh}^2\left( 1+\frac{1}{k_{sh}} \right) \right]}   
\end{eqnarray}
and hence
\begin{equation}
\omega  = 1 +\frac{1}{k_{total}}\langle n \rangle,
\end{equation}
where 
\begin{equation} 
\label{ktotal}
\frac{1}{k_{total}} = \frac{\lambda\alpha \langle n \rangle_s^2 \left( 1 + \frac{1}{k_{s}} \right) +
\lambda(1-\alpha) \langle n \rangle_{sh}^2 \left( 1 + \frac{1}{k_{sh}} \right)}{\langle n \rangle^2} - 1.
\end{equation}
In the limit $\langle n \rangle \to \infty$, we finally obtain:
\begin{equation}
\omega \approx \frac{1}{k_{total}}\langle n \rangle.
\label{omegaf}
\end{equation}
which is precisely (\ref{scav}) with $1/k_{total}$ being the positive constant. Hence the validity of KNO scaling requires that $1/k_{total}$ remains  energy independent and in our analysis of data we will look at this quantity. 

It is possible to improve the quantitative study KNO scaling by analyzing the energy dependence of the C-moments 
($C_q = \langle n^q \rangle / \langle n \rangle^q $). In the limit $\langle n \rangle \to \infty$ they should become energy independent. 
In the case of single NBD models one can derive analytic expressions for the $C_q$ moments as a function of $\langle n \rangle$. 
In \cite{Germano:2021brq} it was shown that 
$$
C_2 = \frac{1}{\langle n \rangle} \, + \, 1 \, + \, \frac{1}{k}
$$
In the large mean multiplicity (and hence energy) limit $C_2$ is constant if $k$ is constant. In \cite{Germano:2021brq} it was shown that 
when $|\eta| < 0.5$ this is approximately the case and $C_2$ is constant. For larger pseudorapidity windows, this is no longer true. The 
higher moments show a more complicated dependence on $\langle n \rangle$ and the data show larger error bars. We will not perform the 
study of the moments in this work. Nevertheless, we we would like to point out that for the  double NBD approach it is possible to establish a
connection between the moments and $k_{total}$. For example, it is easy to show that in the scaling limit we have
$$
C_2 =  \frac{1}{k_{total}}
$$
Before closing this section, we would like to mention that studying KNO scaling belongs to the "bottom-up" approaches to multiplicity 
distributions mentioned in the Introduction and it is not the only scaling that has been investigated. Other empirical regularities have been found in the data, which might have 
consequences for the multiplicity distributions, such as geometric scaling  \cite{Moriggi:2020zbv} (and references therein) 
and entropy scaling \cite{Moriggi:2025qfs}.

\section{Results and discussion}

\begin{figure}[h!]
\centering
\resizebox{\textwidth}{!}{
\begin{tabular}{ccc}
\includegraphics[height=4.0cm]{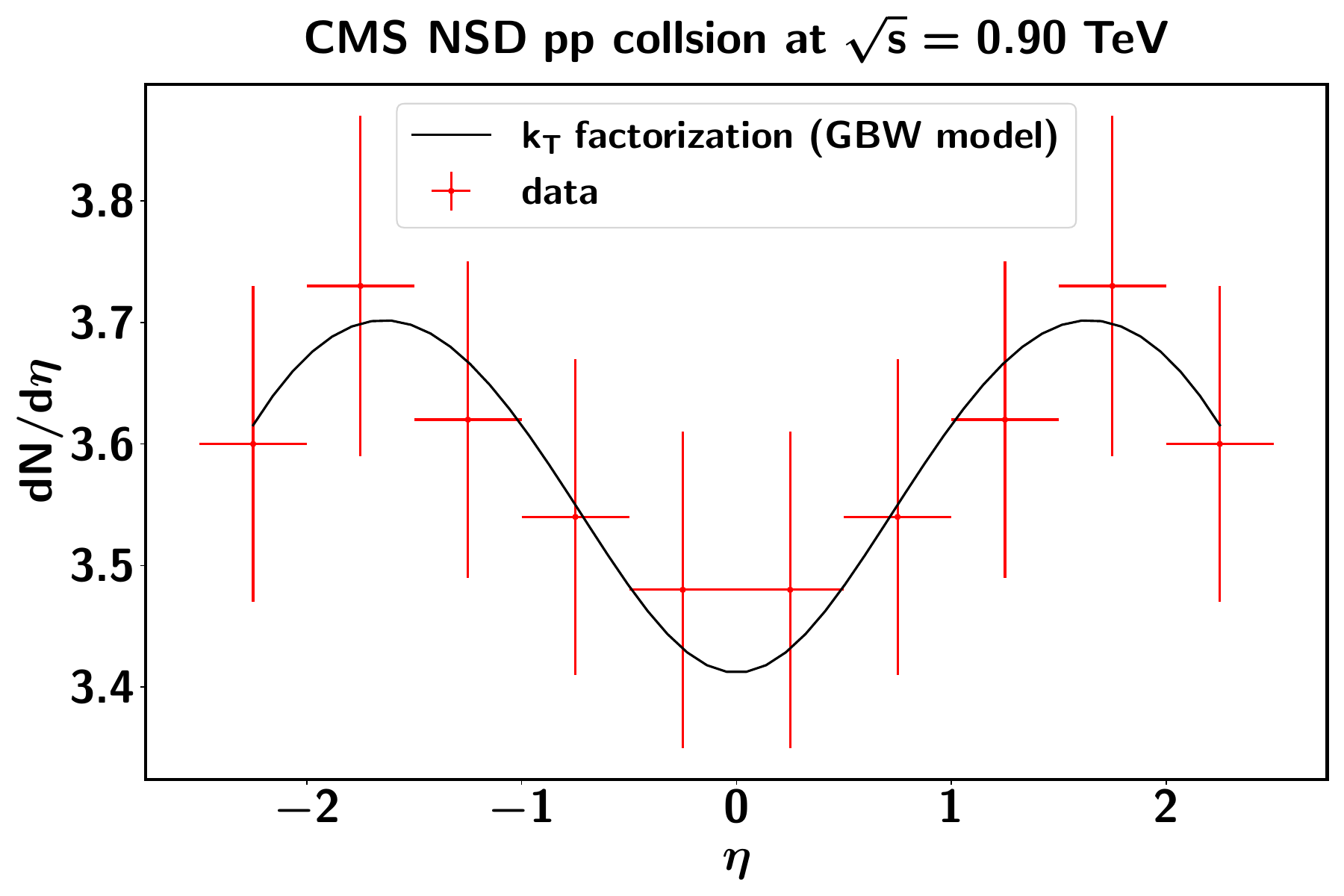} &
\includegraphics[height=4.0cm]{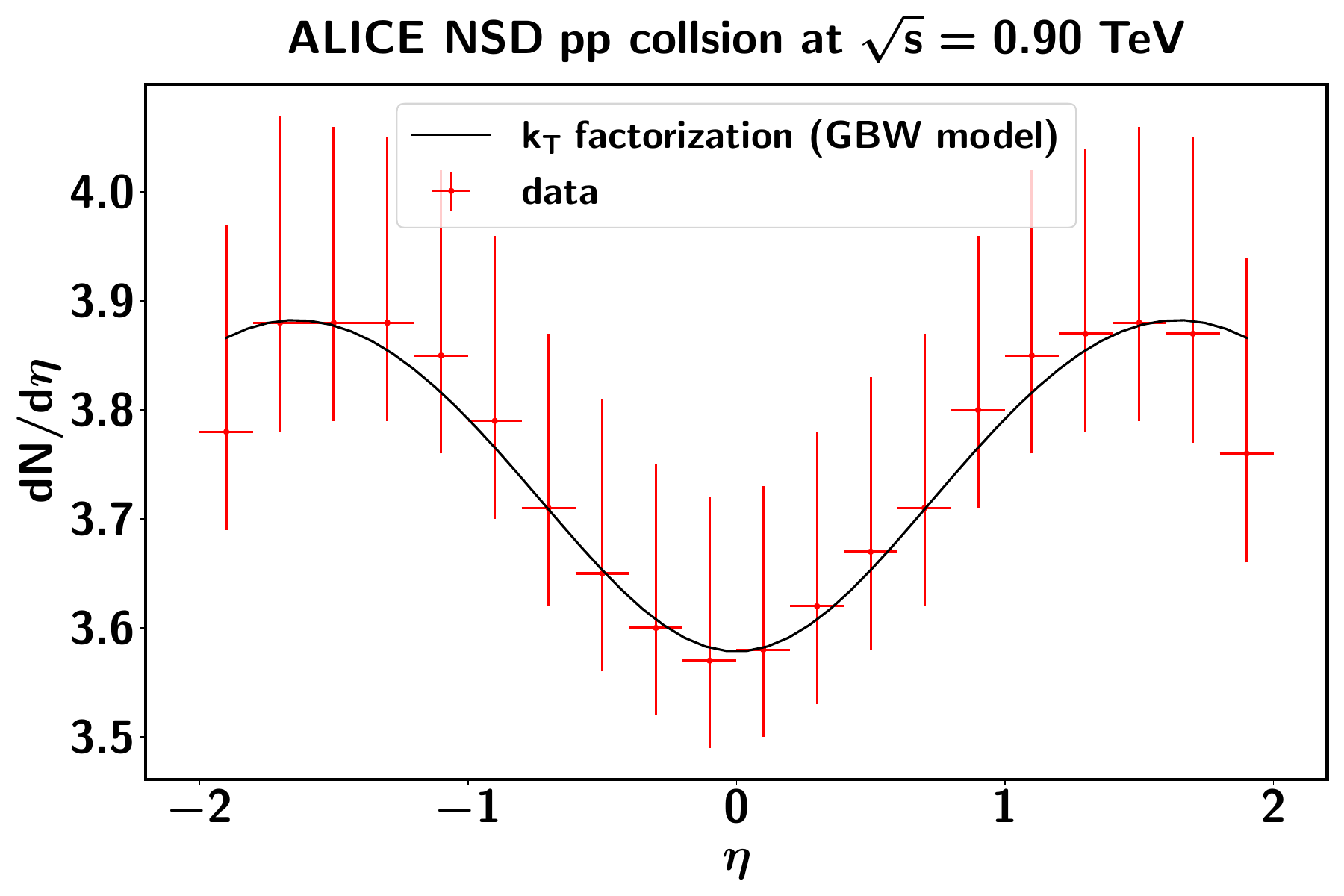} &
\includegraphics[height=4.0cm]{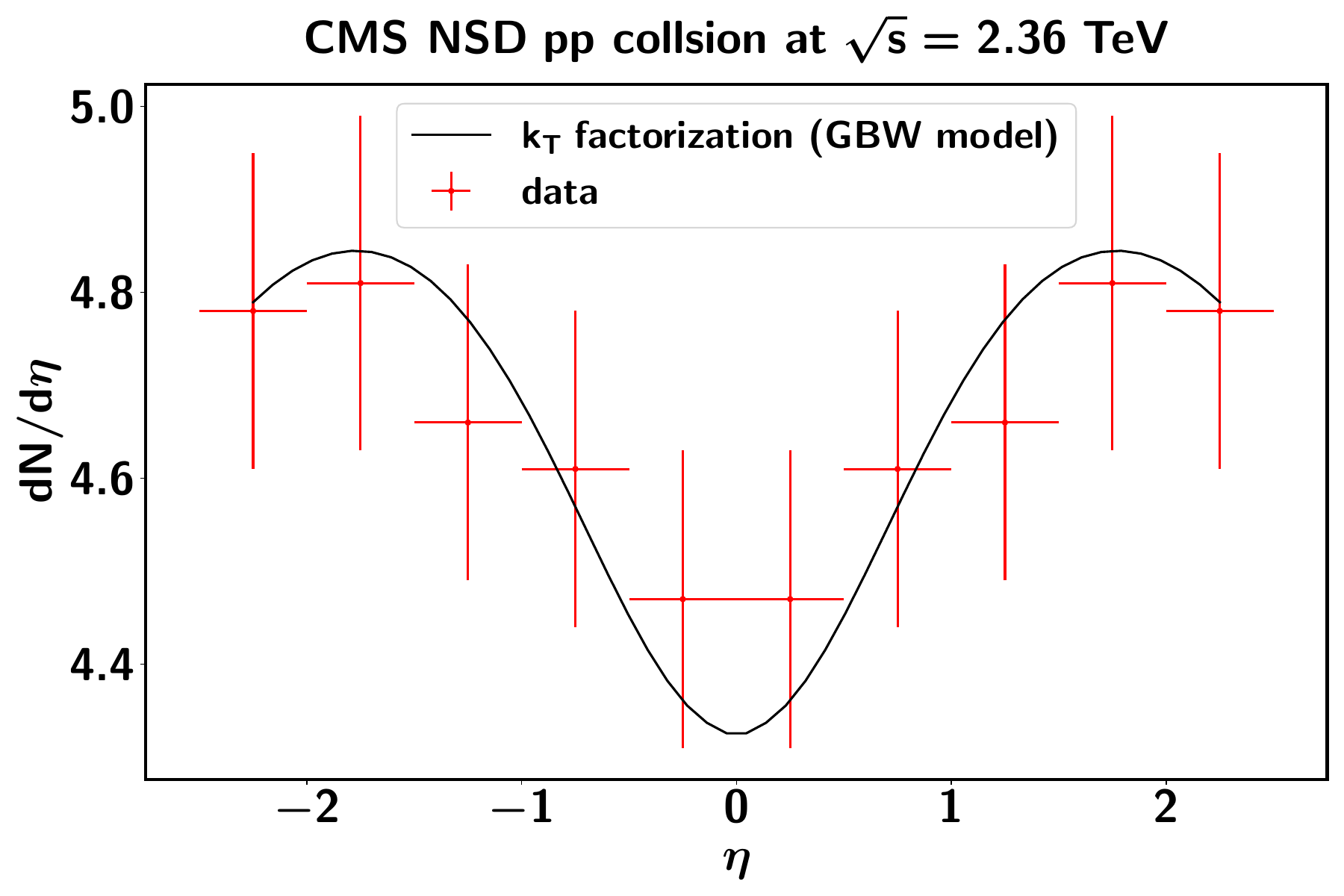} \\
\includegraphics[height=4.0cm]{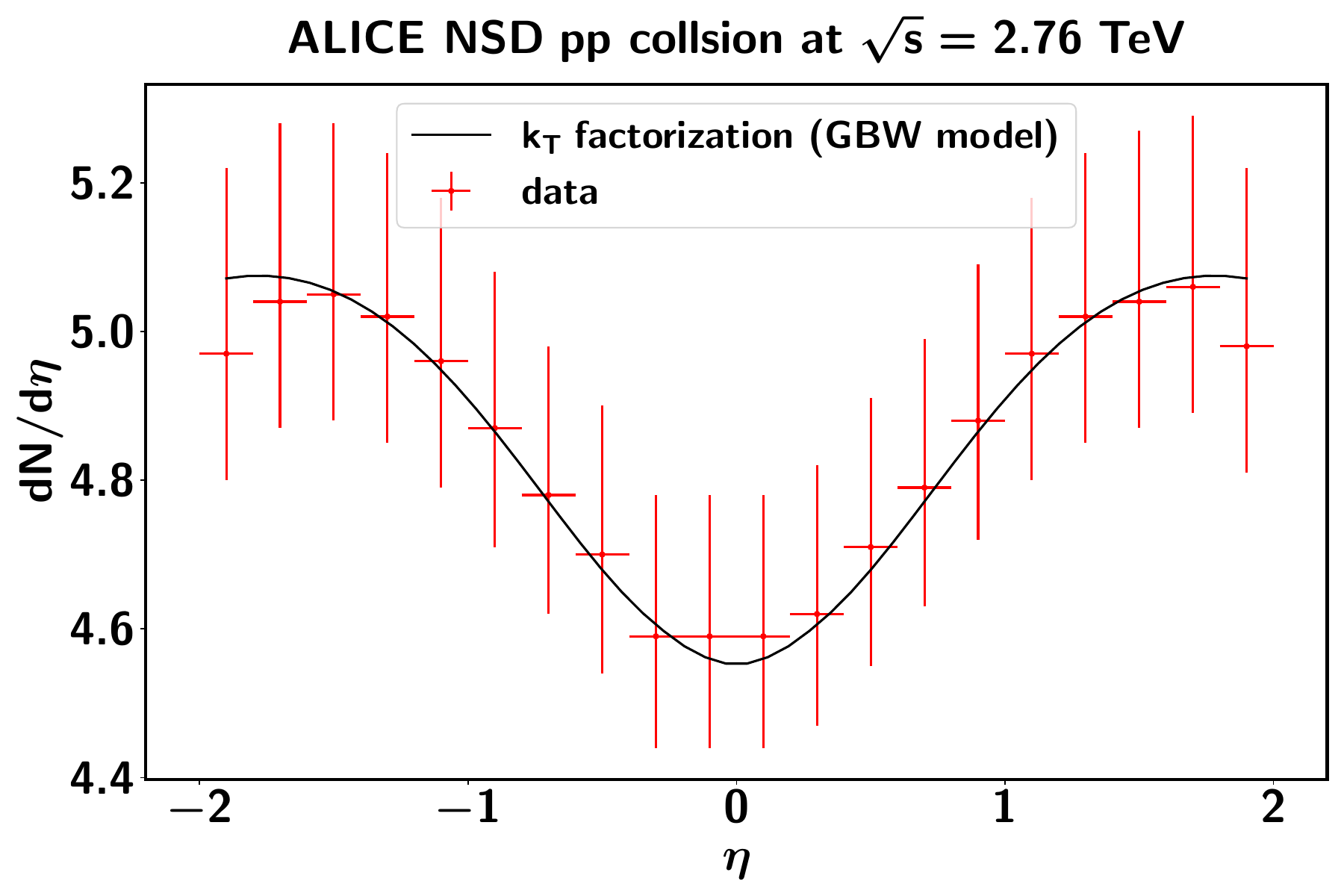} &
\includegraphics[height=4.0cm]{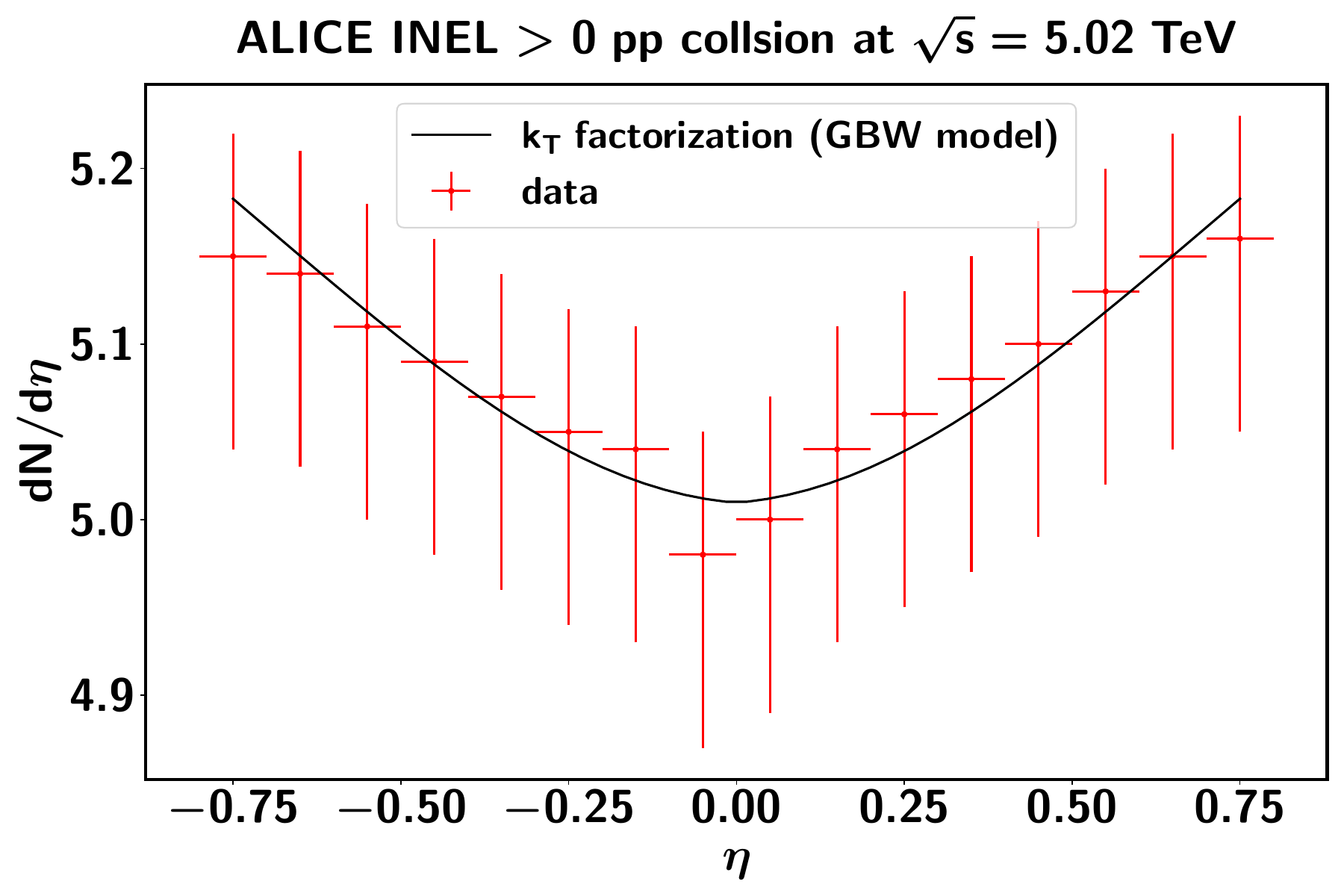} &
\includegraphics[height=4.0cm]{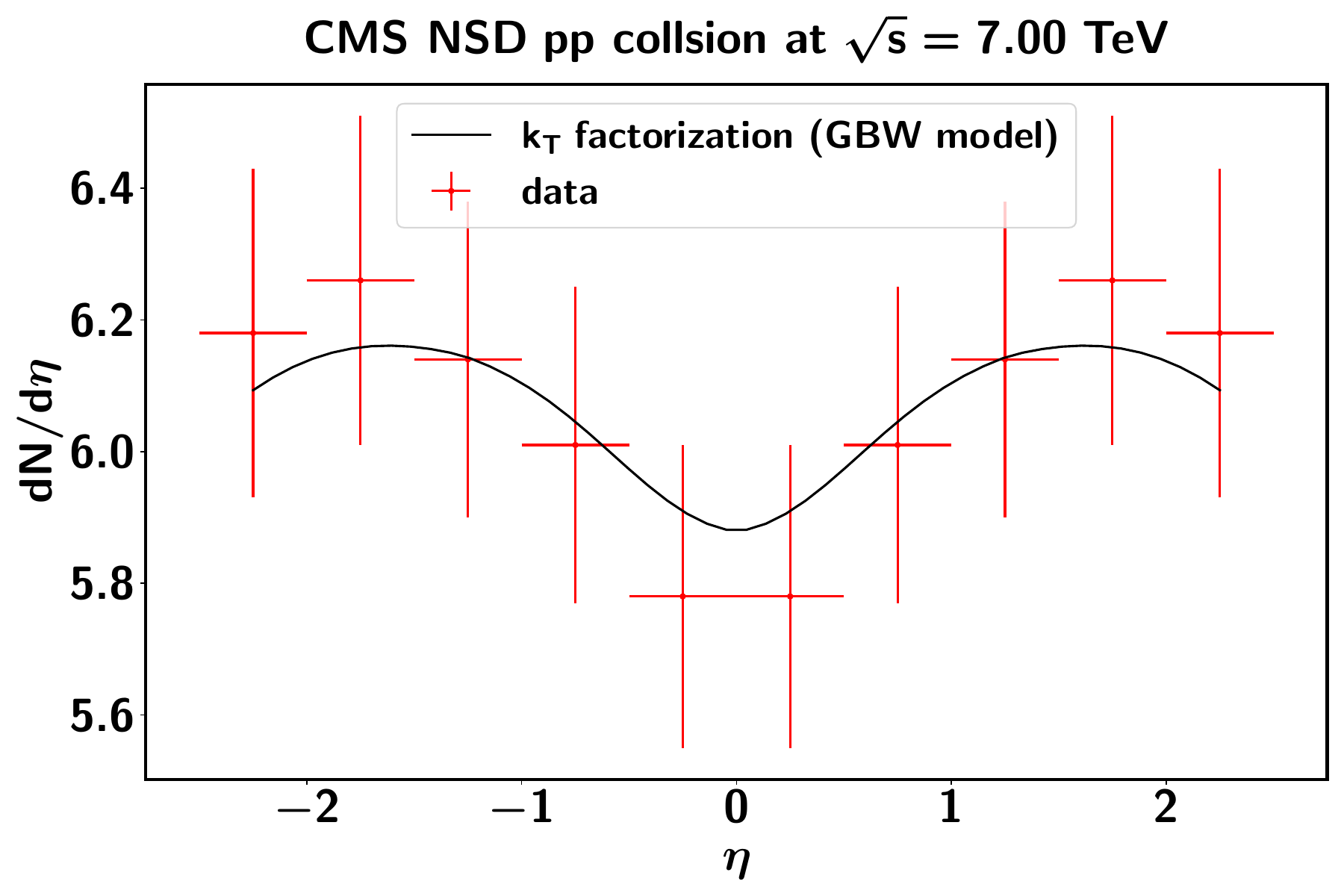} \\
\includegraphics[height=4.0cm]{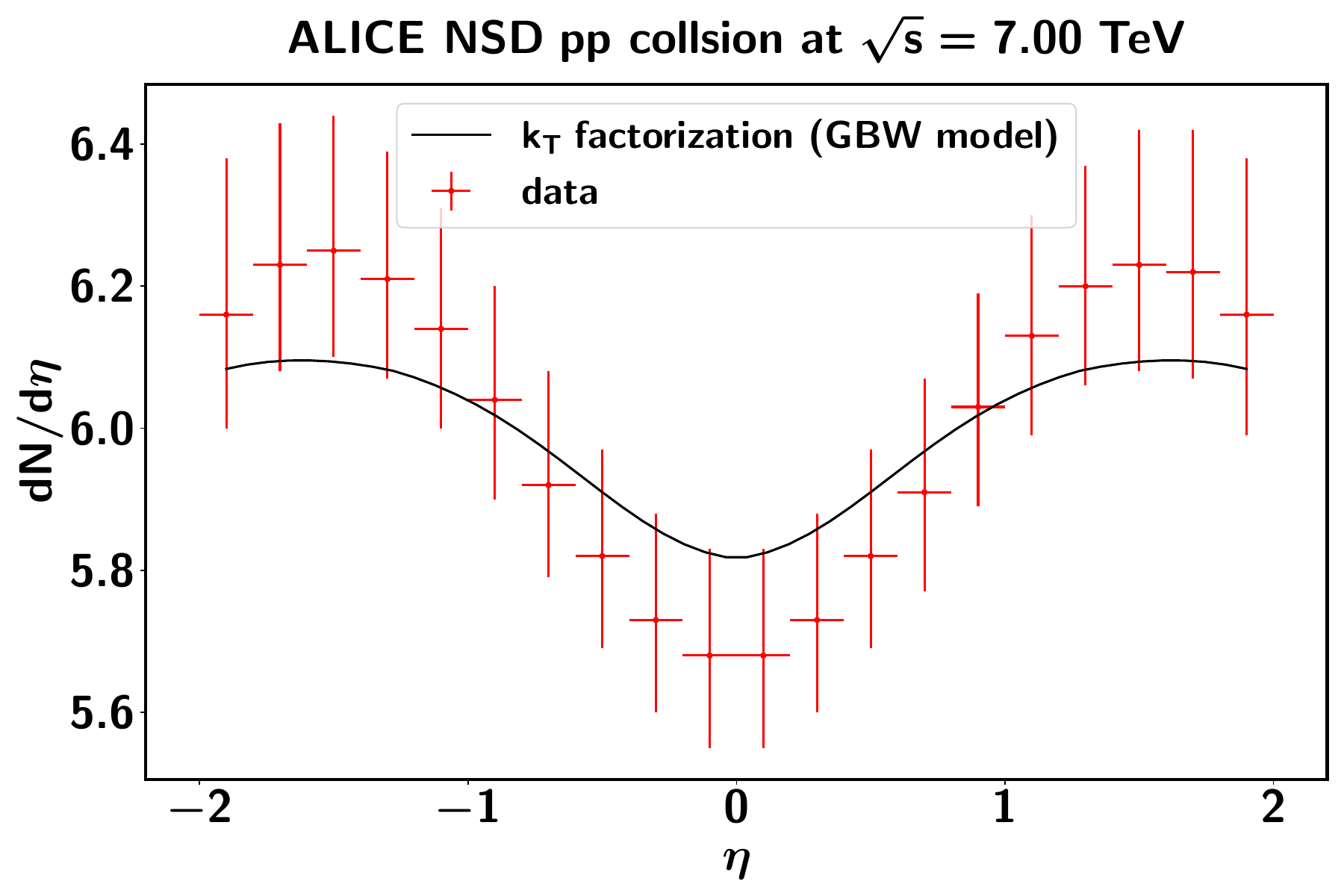} &
\includegraphics[height=4.0cm]{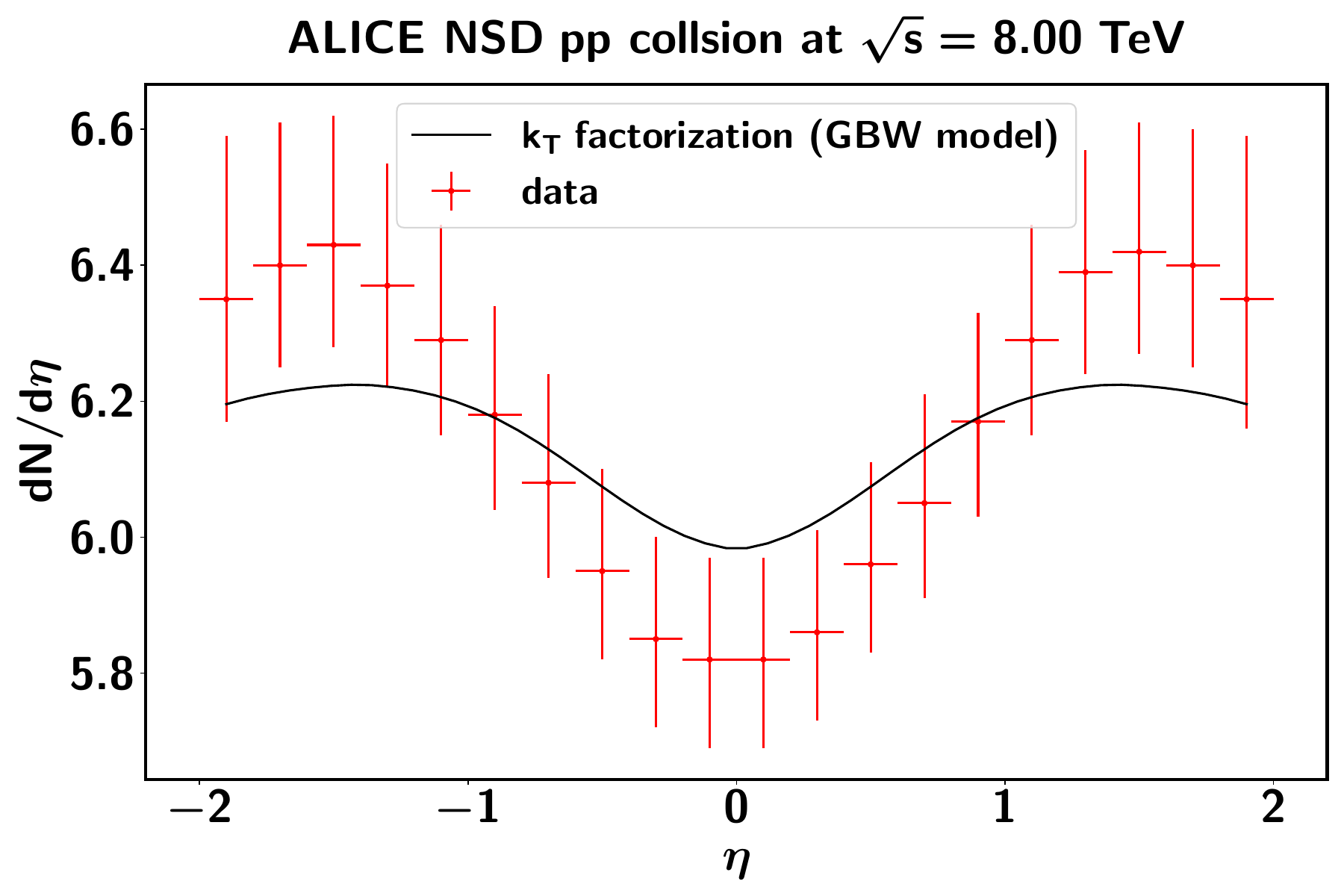} &
\includegraphics[height=4.0cm]{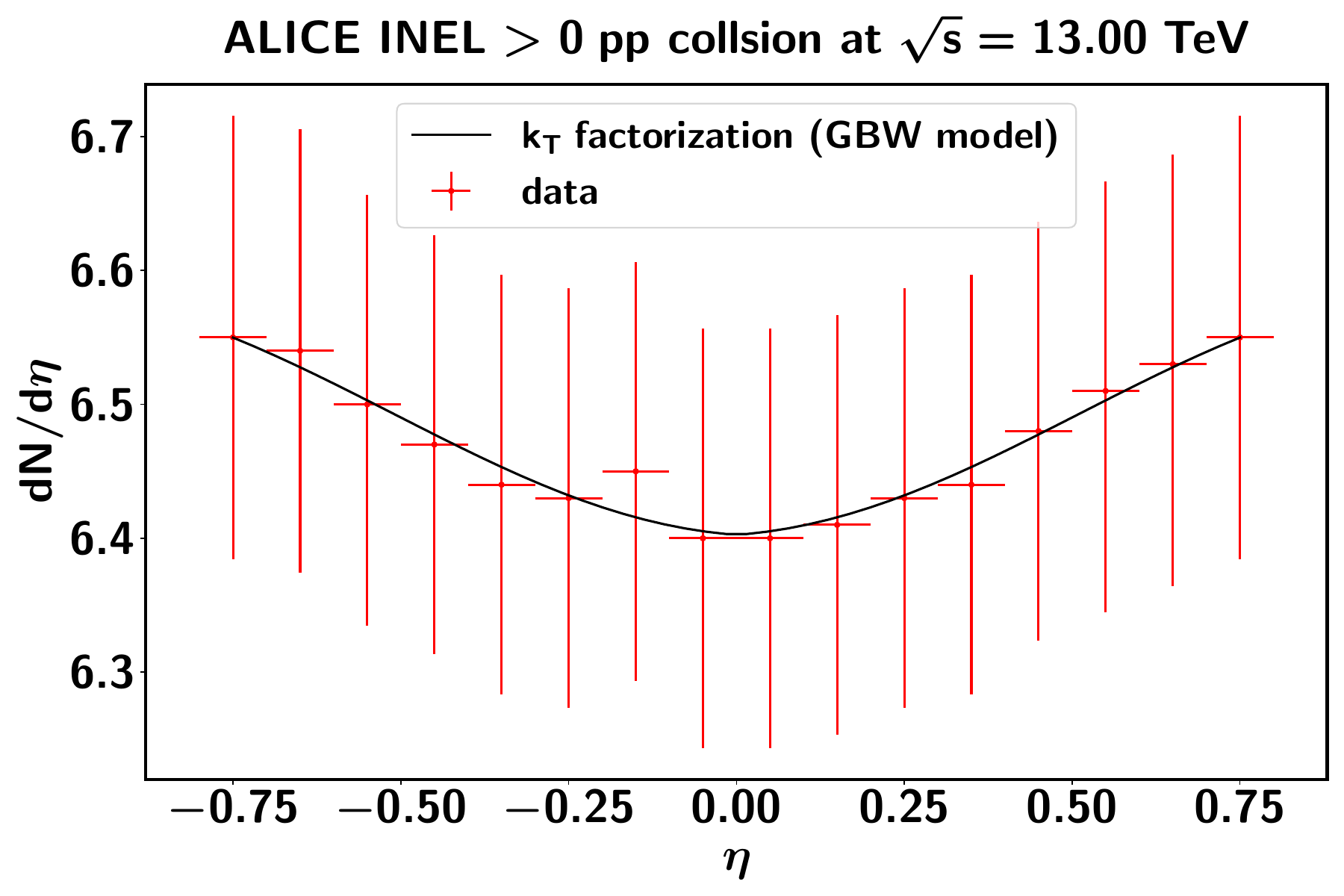} \\
\end{tabular}}
\caption{Pseudorapidity distributions.                            
The solid lines are obtained from Eq. (\ref{sep_final}). Data are 
from CMS \cite{CMS:2010wcx,CMS:2010tjh} and ALICE 
\cite{ALICE:2015olq,ALICE:2022kol}; the bars show the combined uncertainties.}
\label{fig1}
\end{figure}

The global normalization parameter $K/S$ is obtained by fitting Eq. (\ref{sep_final}) to the data at different energies. 
The values of $K/S$ are shown in Table \ref{tab:KS_values}. It is reassuring to see that this normalization factor depends very weakly
on the energy. As can be seen in Fig. \ref{fig1}, we obtain an overall good description of the pseudorapidity distribution data.

\begin{table}[h!]
\centering
\setlength{\tabcolsep}{12pt} 
\begin{tabular}{ccc}
\hline
data &\(\sqrt{s}\) (TeV) & \(K/S\) (GeV$^2$) \\
\hline
\cite{CMS:2010wcx}&0.9   &  0.0968 $\pm$ 0.0011\\
\cite{ALICE:2015olq}&0.9   &  0.1015 $\pm$ 0.0011\\
\cite{CMS:2010wcx}&2.36  &  0.0993 $\pm$ 0.0011\\
\cite{ALICE:2015olq}&2.76   & 0.1001 $\pm$ 0.0012\\
\cite{ALICE:2022kol}&5.02   &  0.0929 $\pm$ 0.0005\\
\cite{CMS:2010tjh}&7.0   &  0.0975 $\pm$ 0.0012\\
\cite{ALICE:2015olq}&7.0   &  0.0965 $\pm$ 0.0007\\
\cite{ALICE:2015olq}&8.0  & 0.0952 $\pm$ 0.0007\\
\cite{ALICE:2022kol}&13.0  &  0.0888 $\pm$ 0.0005\\
\hline
\end{tabular}
\caption{Values of \( K/S \) for different energies. The data used for the fits are 
from CMS \cite{CMS:2010wcx,CMS:2010tjh} and ALICE \cite{ALICE:2015olq,ALICE:2022kol}.}
\label{tab:KS_values}
\end{table}

To obtain the soft and semihard average multiplicities from (\ref{sep_final}) we will fix the separation scale to be $\Lambda = 1.4$  GeV.
This is approximately the same number used in \cite{Broilo:2019yuo}  and is in the range of usual values used to define the separation between soft and semihard physics (see  \cite{Iser:2025zhz} and \cite{Broilo:2019yuo} for further references). 
Having fixed these numbers, we use equation (\ref{sis}) together with (\ref{DNBD}) to perform a fit of all the LHC data on MDs in 
order to determine the parameters $\lambda$, $\alpha$, $k_s$ and $k_{sh}$. The values of these parameters can be found in Table \ref{tab-NBD} in the Appendix, together with the values obtained for the multiplicities and for $k_{total}$.   

From  Table \ref{tab-NBD} we can infer that
the variables which are usually called $\langle n_1 \rangle$ and $ \langle n_2 \rangle$ indeed correspond to $\langle n_s \rangle$ and 
$\langle n_{sh} \rangle$  respectively. Moreover the previously found relation
$\langle n_2 \rangle \simeq 3 \langle n_1 \rangle$  becomes $\langle n_{sh} \rangle \simeq 3 \langle n_s \rangle$ and is still valid, 
changing slightly with the center-of-mass energy and with the 
size of the  rapidity window. This is consistent with the findings of the ALICE  study published in \cite{ALICE:2015olq} and in the analysis of CMS data perfomed in  \cite{Ghosh:2012xh}. 

\begin{figure}[!t]
\centering
\resizebox{\textwidth}{!}{
\begin{tabular}{ccc}
\includegraphics[height=4.0cm]{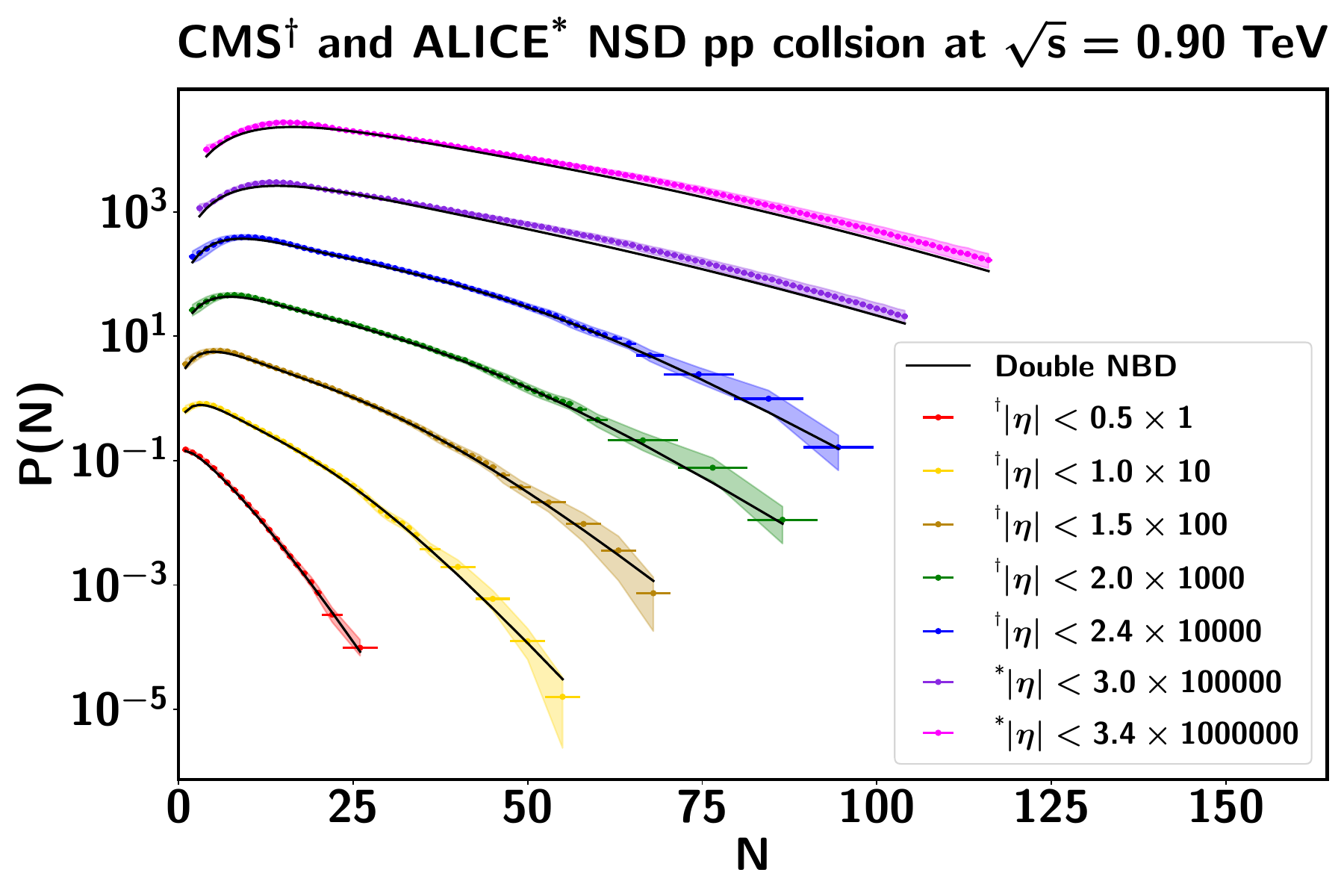} &
\includegraphics[height=4.0cm]{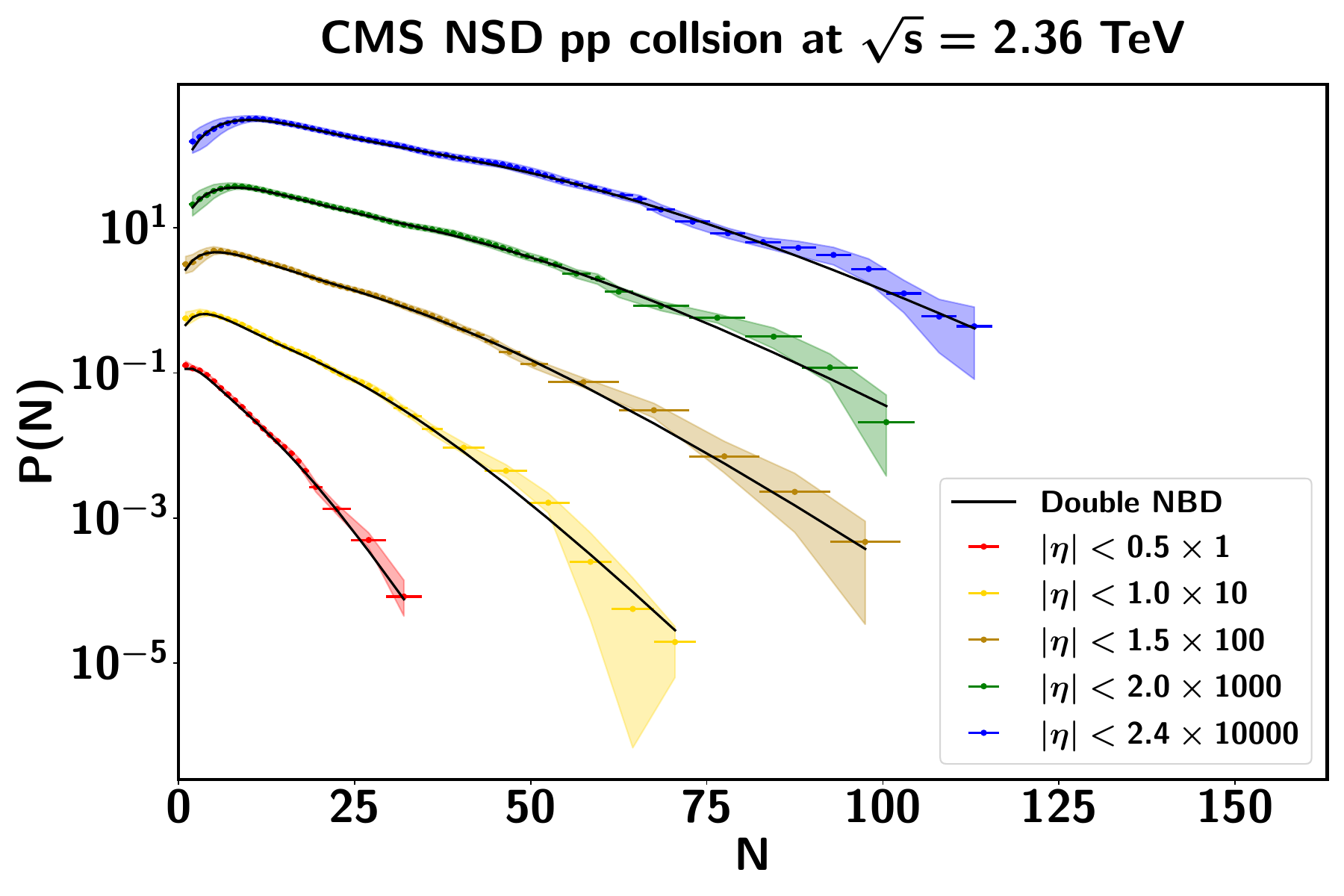} &
\includegraphics[height=4.0cm]{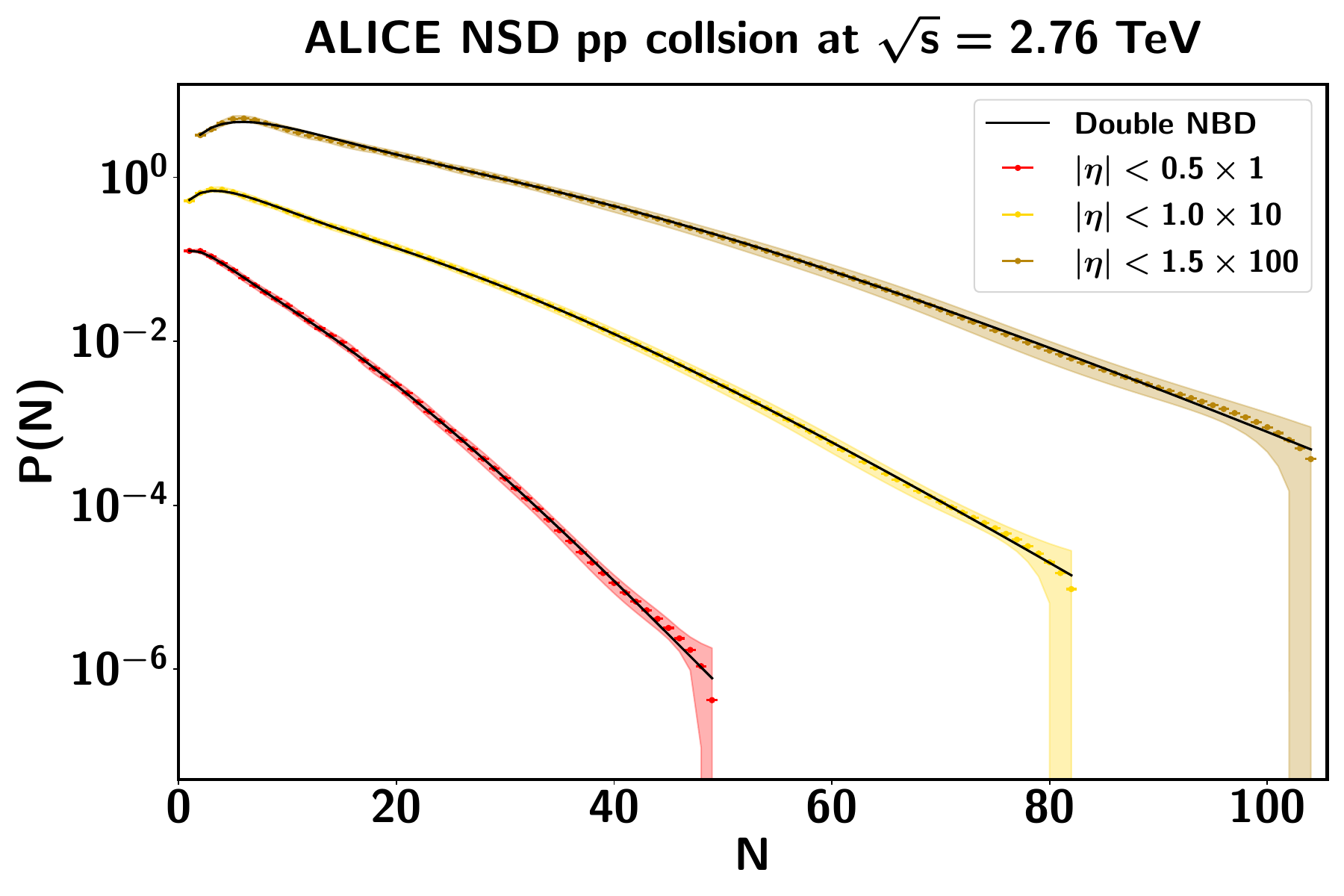} \\
\includegraphics[height=4.0cm]{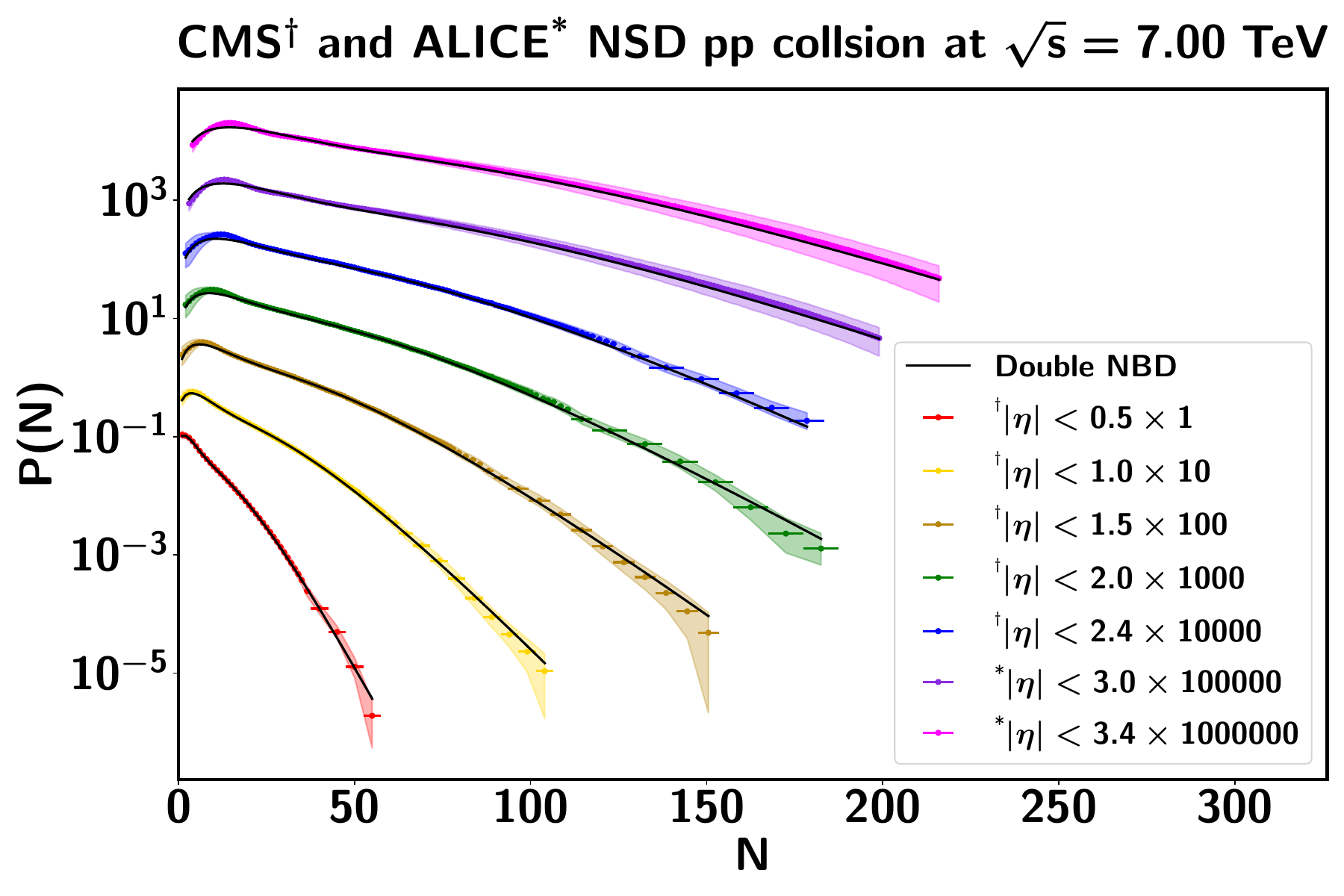} &
\includegraphics[height=4.0cm]{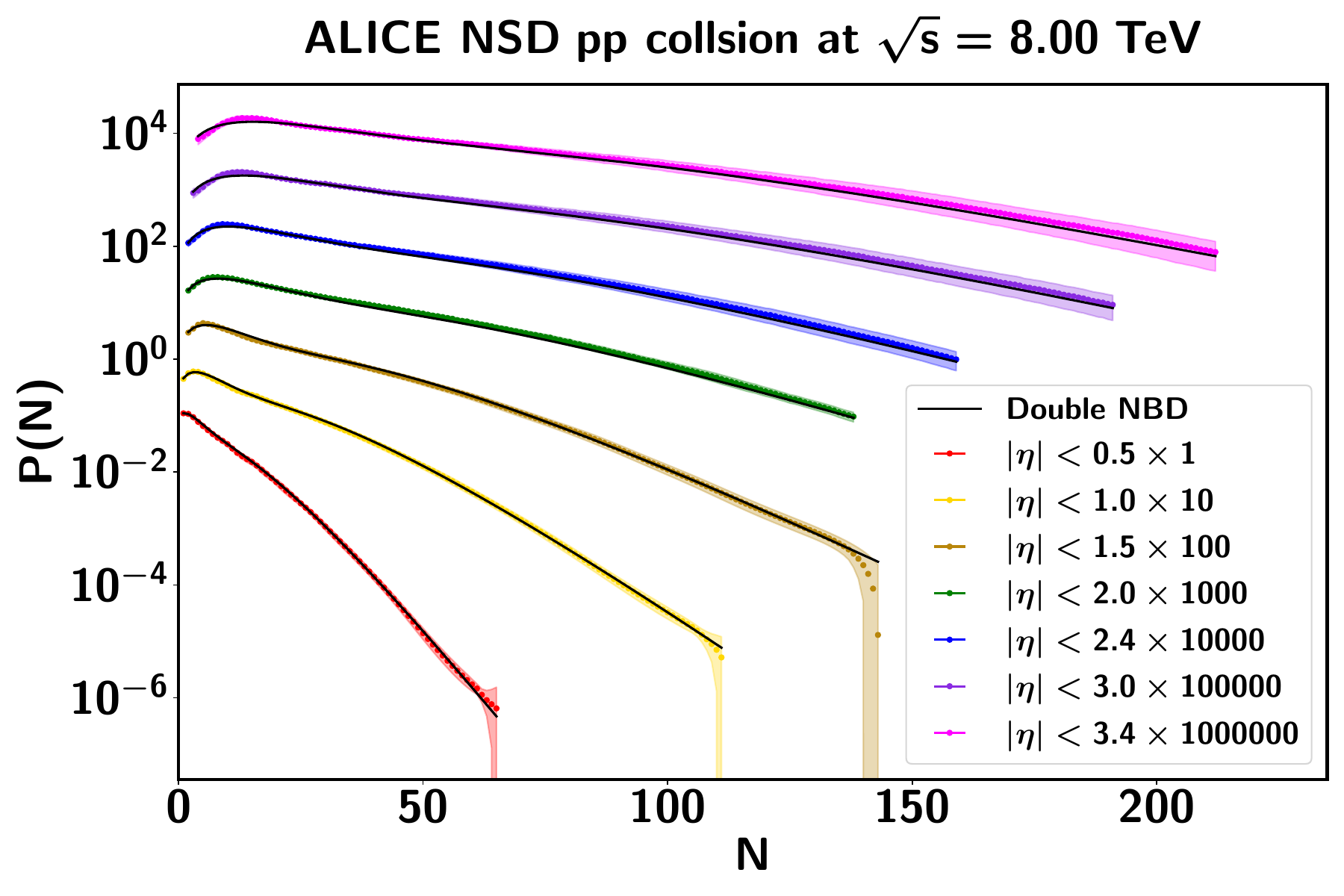} &
\includegraphics[height=4.0cm]{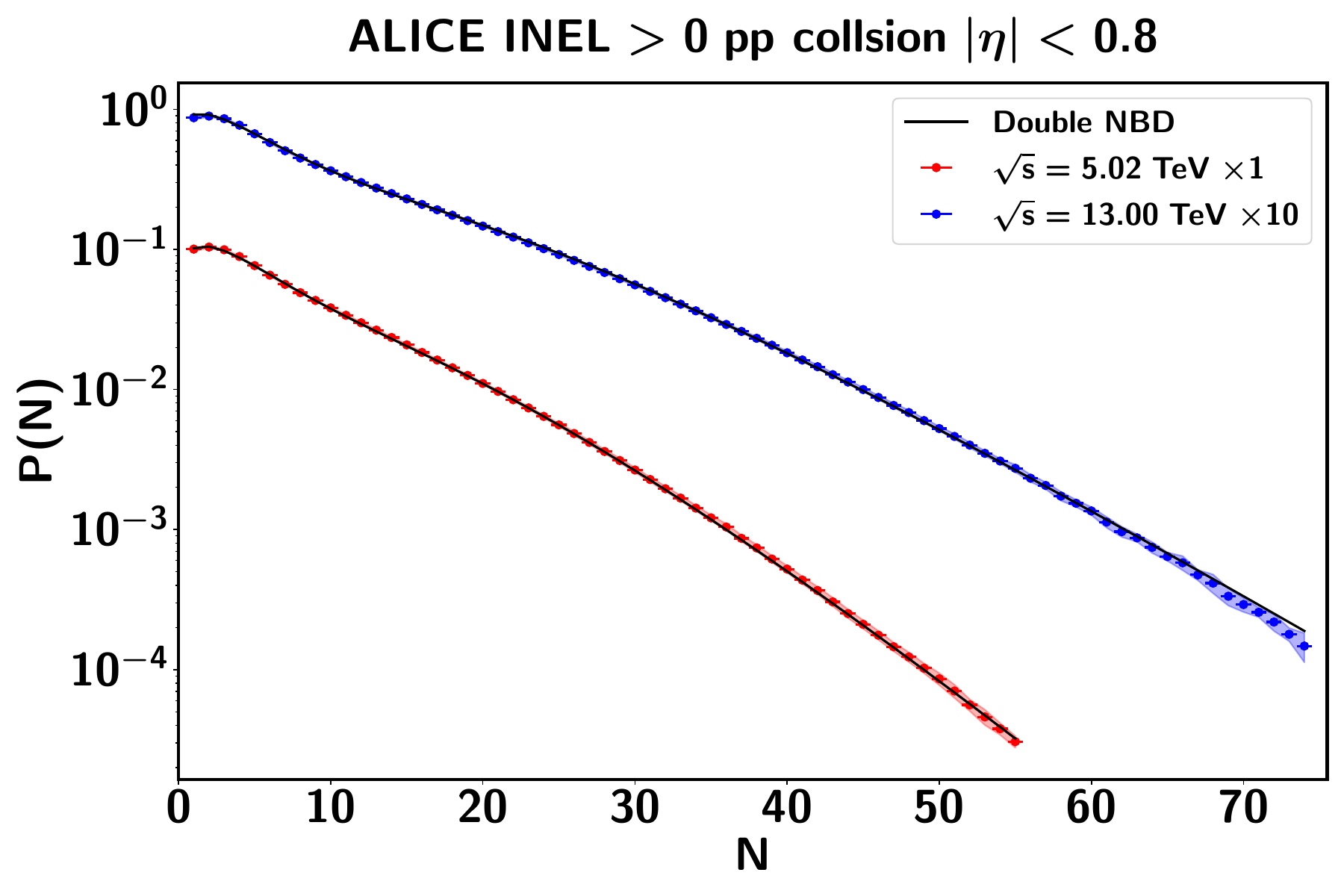} \\
\end{tabular}}
\caption{Multiplicity distributions measured by different collaborations.  In the bottom-right panel we show the MDs measured in inelastic events (INEL). All the other panels show data from non-single-diffractive (NSD) events. The fits are obtained with Eq. (\ref{DNBD}) and the fit parameters are
listed in Table \ref{tab-NBD}. Data are from CMS \cite{khachatryan2011charged} and ALICE \cite{ALICE:2015olq,acharya2017charged,acharya2023multiplicity}. The colored bands show the combined uncertainties.}
\label{fig2}
\end{figure}

In Fig. \ref{fig2}, we show the comparison between the data and our model.  We obtain fits of very good quality, which is reflected in the low values of $\chi^2$, among the smallest values reported in the literature for this class of model. The DNBD function provides a precise description of the entire set of multiplicity distributions  measured at the LHC. This result is particularly remarkable given the reduction of free parameters from 6 to 4.

\begin{figure}[h!]
 \centering
 \begin{subfigure}[b]{0.45\textwidth}
    \includegraphics[width=\linewidth]{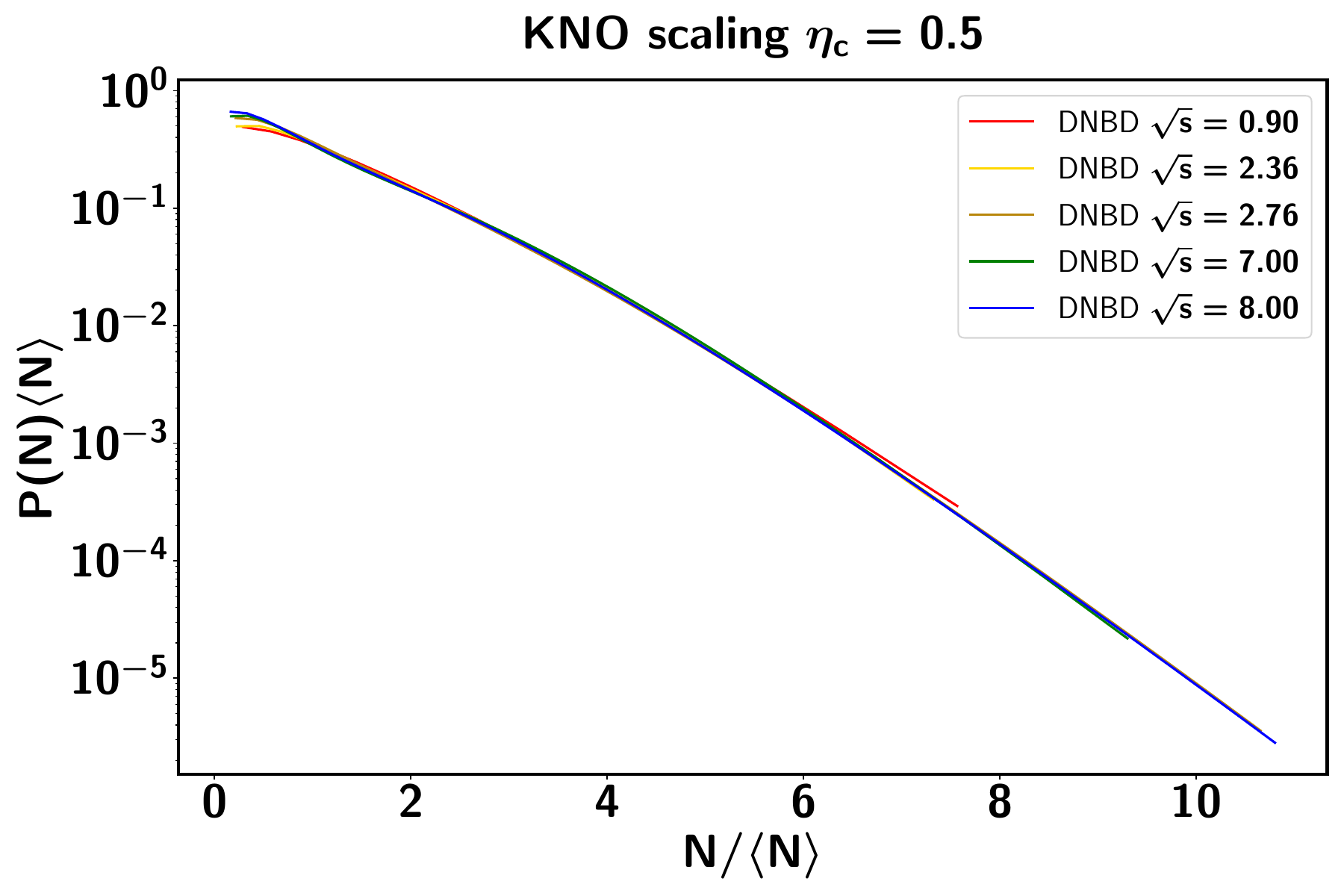}
    \caption{} \label{fig:kno05}
  \end{subfigure}
  \hfill
  \begin{subfigure}[b]{0.45\textwidth}
    \includegraphics[width=\linewidth]{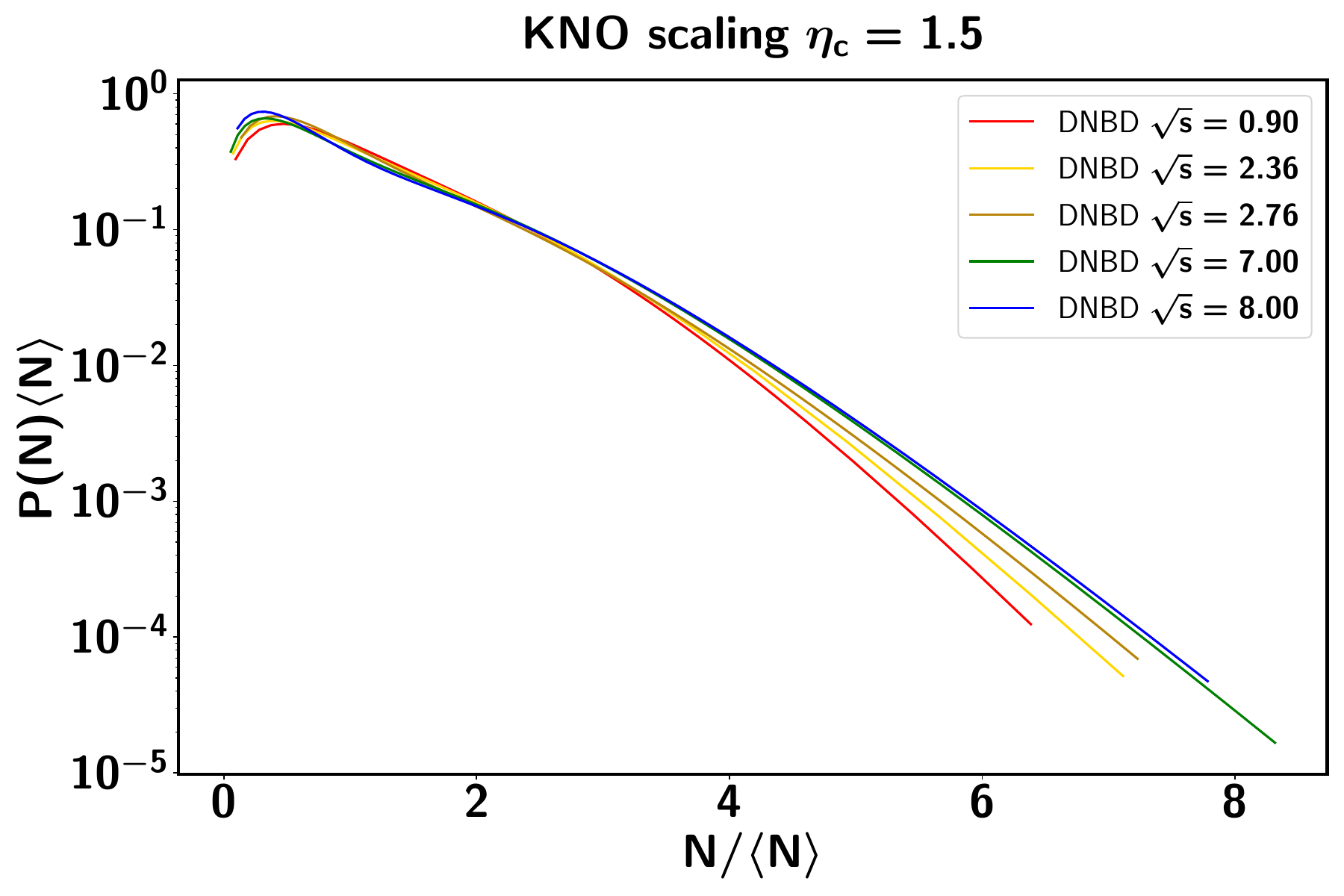}
    \caption{} \label{fig:kno15}
  \end{subfigure}
  \vspace{1em}
  \begin{subfigure}[b]{0.45\textwidth}
    \includegraphics[width=\linewidth]{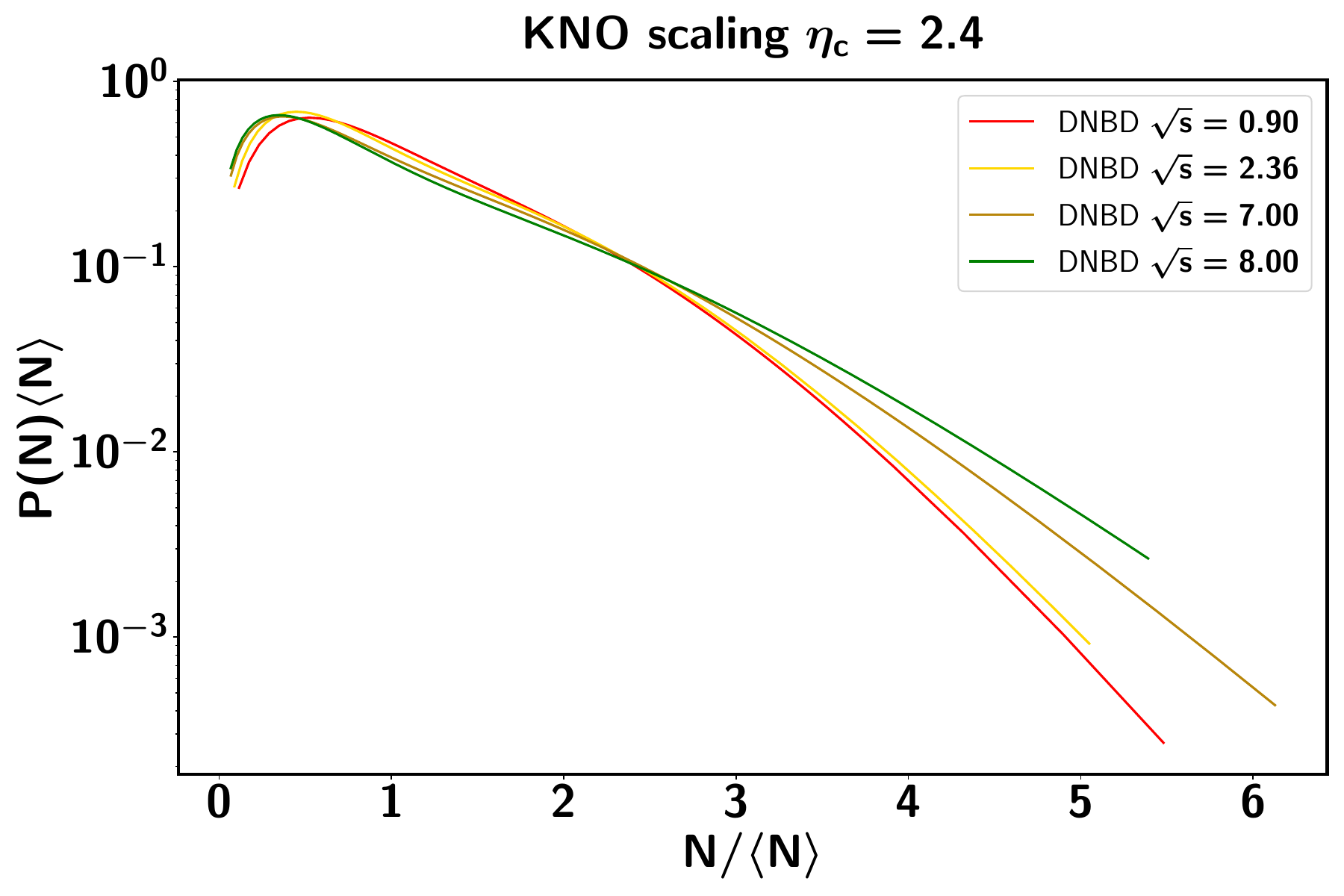}
    \caption{} \label{fig:kno24}
  \end{subfigure}
  \hfill
  \begin{subfigure}[b]{0.45\textwidth}
    \includegraphics[width=\linewidth]{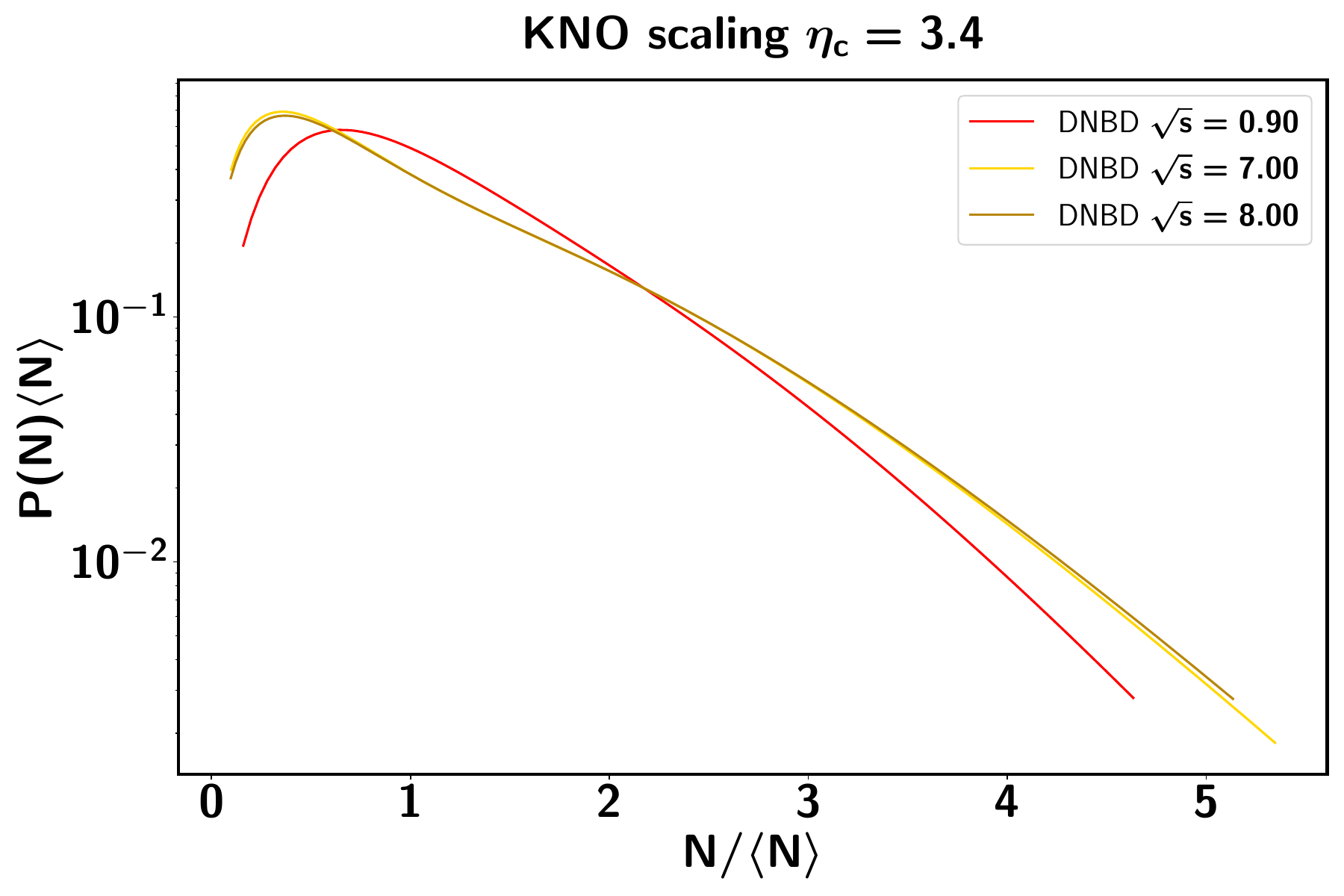}
    \caption{} \label{fig:kno34}
  \end{subfigure}
 \caption{KNO scaling of multiplicity distributions for various pseudorapidity 
windows ($- \eta_c < \eta < + \eta_c$).}
  \label{fig:KNO}
\end{figure}

\begin{figure}[h!]
  \centering
  \begin{subfigure}[b]{0.45\textwidth}
\includegraphics[width=\linewidth]{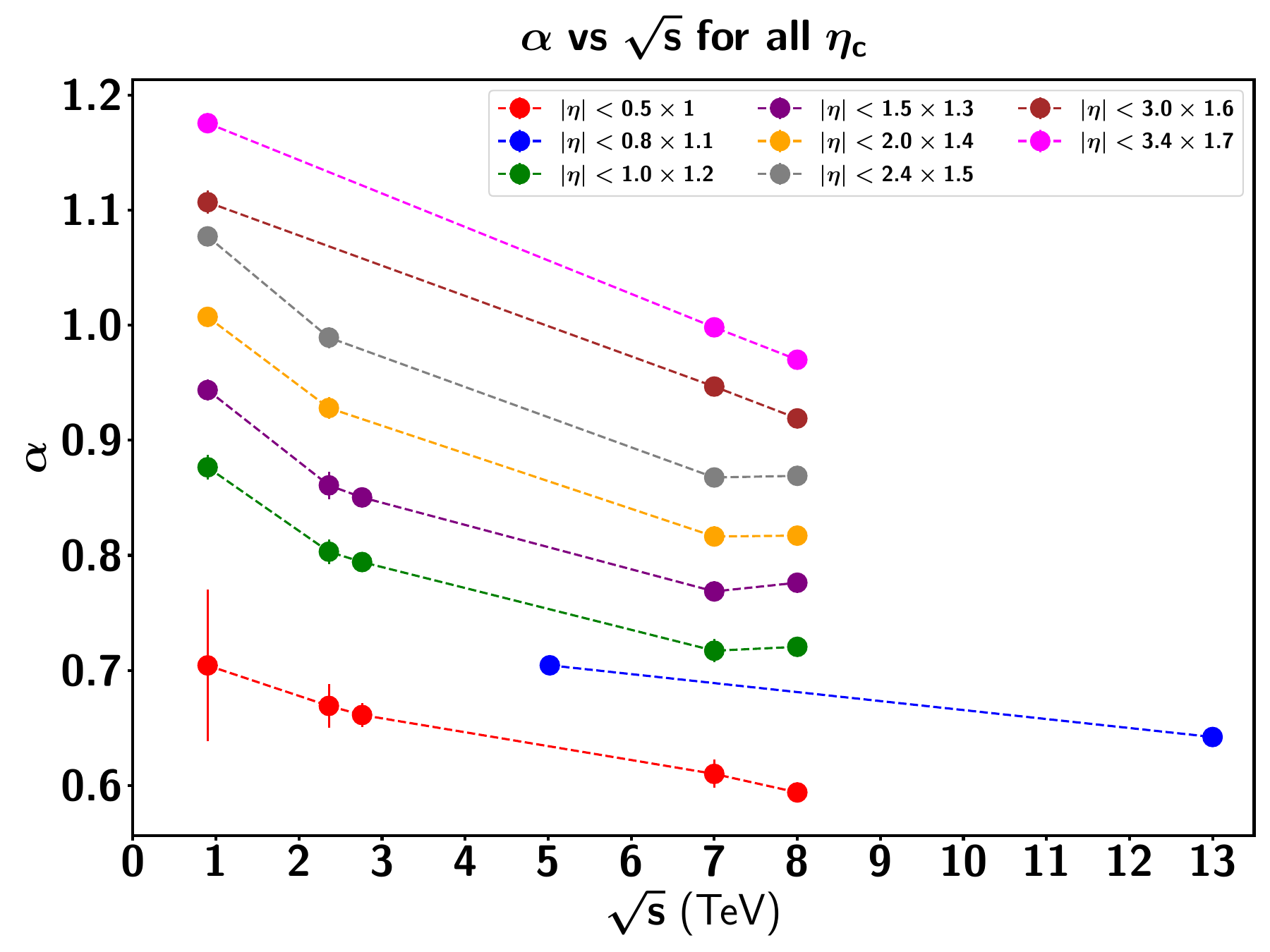}
    \caption{} \label{fig:alpha}
  \end{subfigure}
  \hfill
  \begin{subfigure}[b]{0.45\textwidth}
    \includegraphics[width=\linewidth]{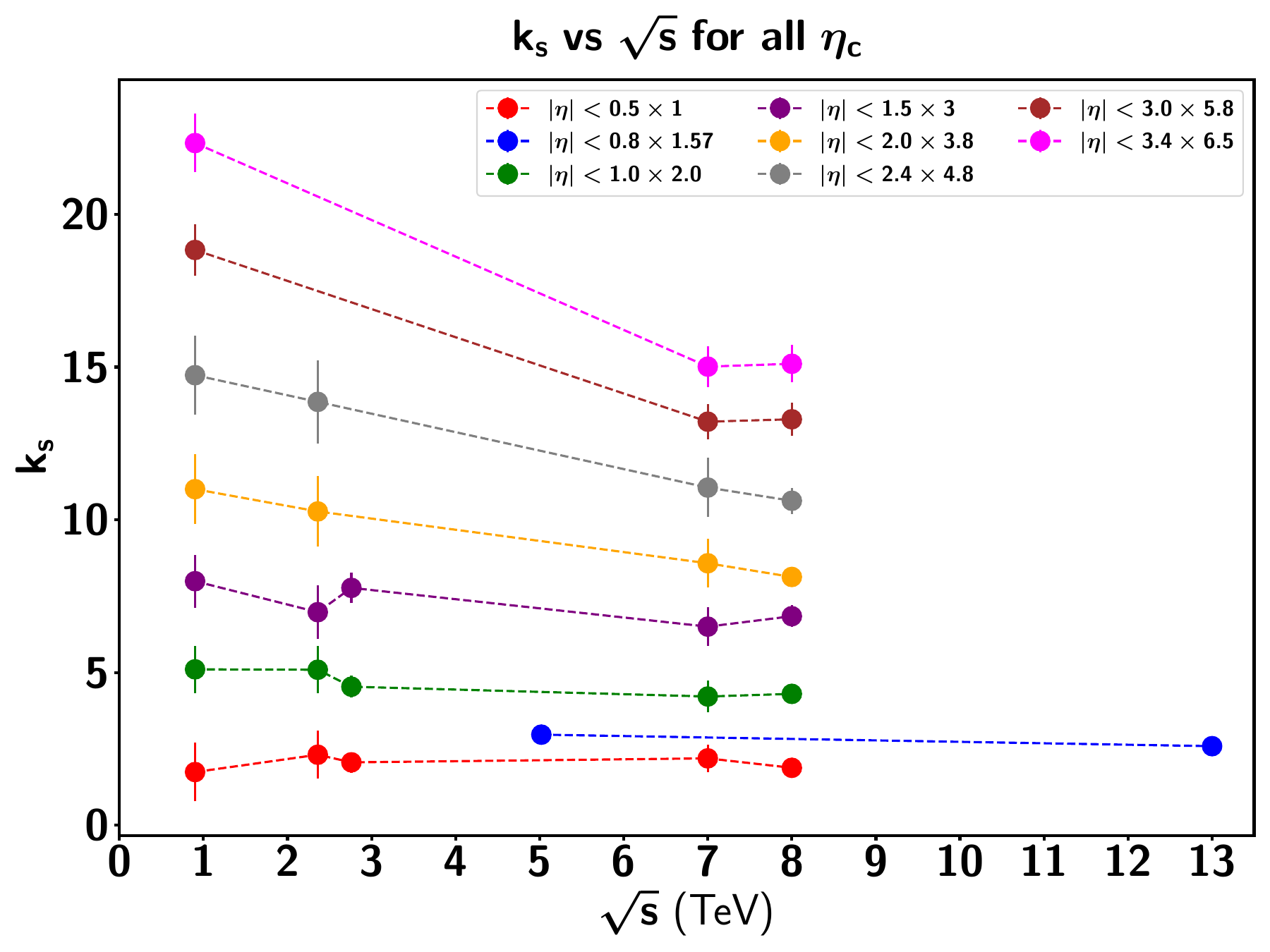}
    \caption{} \label{fig:ks}
  \end{subfigure}

  \vspace{1em}  
  \begin{subfigure}[b]{0.45\textwidth}
    \includegraphics[width=\linewidth]{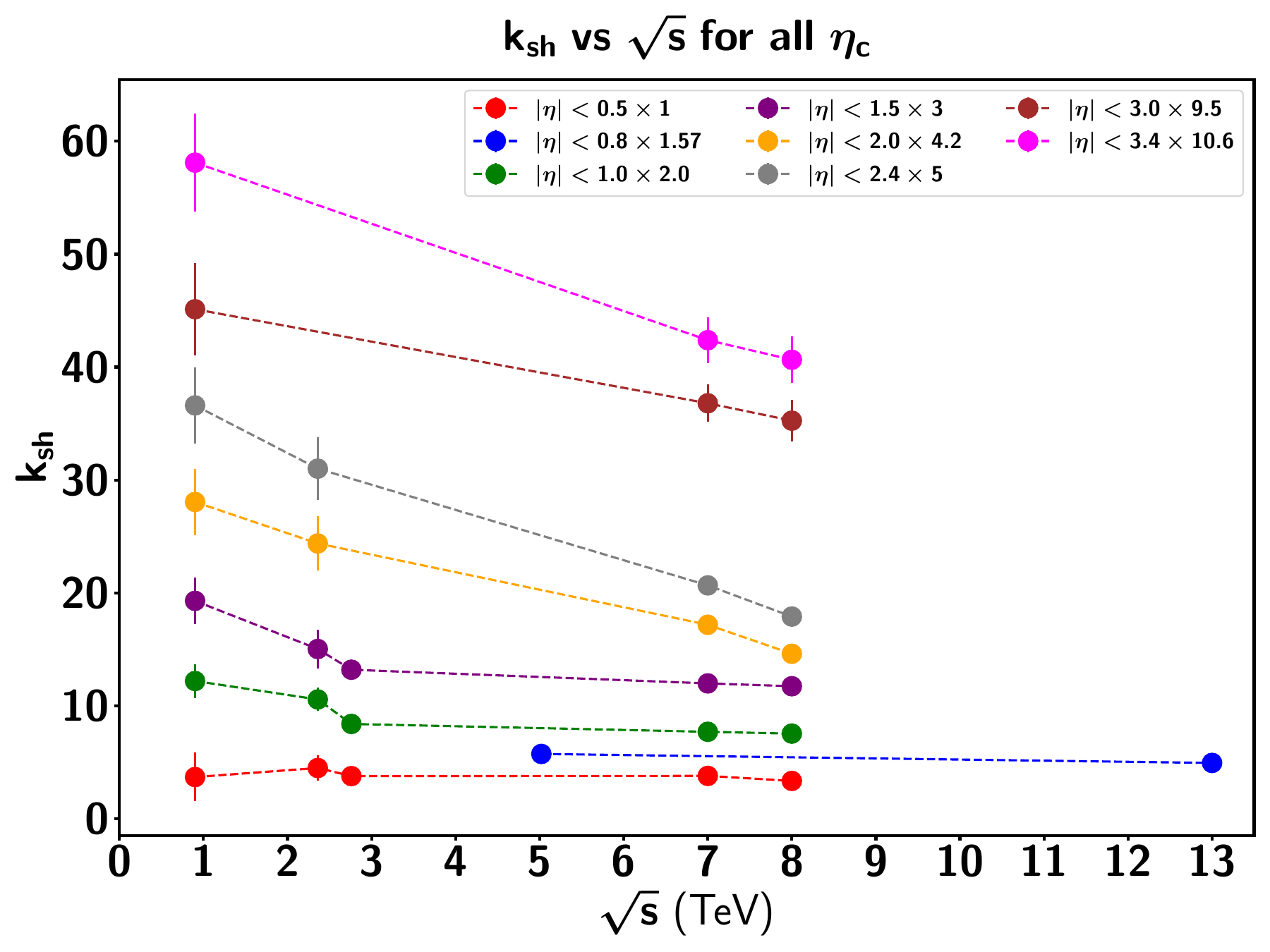}
    \caption{} \label{fig:ksh}
  \end{subfigure}
  \hfill
  \begin{subfigure}[b]{0.45\textwidth}
    \includegraphics[width=\linewidth]{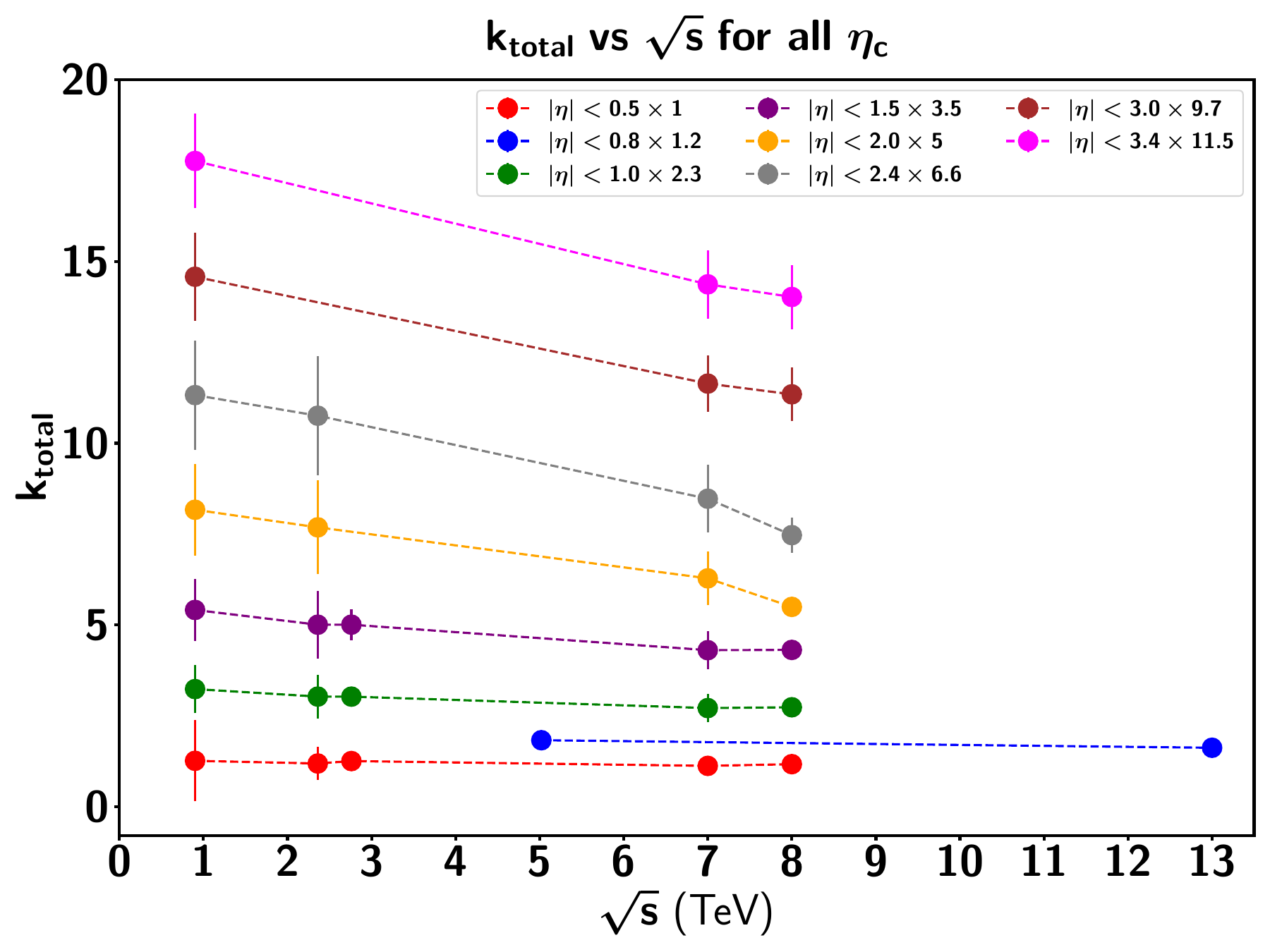}
    \caption{} \label{fig:ktotal}
  \end{subfigure}

  \caption{Behavior of $\alpha$, $k_{s}$, $k_{sh}$ and $k_{\rm total}$ with increasing energy for various pseudorapidity windows ($- \eta_c < \eta < + \eta_c$). The lines are just to guide the eyes. The data points in blue refet to INEL events. All the others refer to NSD events. }
  \label{fig:alpha_ks_fullpage}
\end{figure}

In order to verify whether KNO scaling is observed in the LHC data, we could simply plot the data. However,  this would produce some
visually cumbersome figures, specially when including the errors. Instead, since our fit of the data is very good, we  plot the 
multiplicity distributions computed with Eq. (\ref{DNBD}) and presented in terms of the KNO variable.  The resulting curves are shown 
in Fig. \ref{fig:KNO}. It is very clear that for the smallest pseudorapidity window ($|\eta_c| < 0.5$) KNO scaling holds. For the others, 
the larger the window, the more KNO scaling is violated.  In \cite{Dokshitzer:2025owq} it was shown that in the very high energy limit
pQCD predicts scaling.  Having this in mind,  we might think that the perturbative events are concentrated in the narrow, central rapidity window and are responsible for the scaling, whereas the nonperturbative events are increasingly important at larger rapidity windows. If this were the case, we would expect that the parameter $\alpha$ would fall strongly with the energy for $|\eta_c| < 0.5$ and
stay constant or increase in the other windows. However, this is not the case.   In Fig. \ref{fig:alpha} we show the behavior of $\alpha$ with $\sqrt{s}$ for all windows multiplied by some numerical factor for the sake of clarity. We can see that the falling behavior is the same for 
all windows and, contrary to previous expectations, there is no correlation between $\alpha$ and the violation of KNO scaling. In fact, if
we did not multiply the values of $\alpha$ by the separating numerical factors, they would fall onto a single curve (within a few percent of uncertainty). Interestingly, in 
\cite{ALICE:2015olq} the ALICE Collaboration performing a similar DNBD analysis of the same data found that $\alpha$ increases with the 
energy $\sqrt{s}$ in all rapidity windows (see Table 10 in that reference). This is an example of how the introduction of constraints can
bring about not only quantitative but also qualitative changes in the behavior of the fit parameters.

As for the other parameters, we see a marked
change in the behavior of  $k_s$ and $ k_{sh}$  with $\sqrt{s}$, going from constant when $|\eta_c| < 0.5$ to rapidly falling when
$|\eta_c| > 1.0$, as can be seen in Figs.  \ref{fig:ks} and  \ref{fig:ksh}.
Giovannini and Ugoccioni proposed in \cite{Giovannini:1999tw,Giovannini:1998zb} three scenarios for the behavior of $k_s $ and $ k_{sh}$:
\vskip0.1cm
\noindent
i) Both $k_s $ and $ k_{sh} $ remain constant with energy, which means that KNO scaling is valid for both soft and semi-hard  components.
\vskip0.1cm
\noindent
ii) $k_s $ is constant with energy and $ k_{sh} $ decreases linearly with increasing energy, implying that  KNO scaling is strongly violated by the semihard component. 
\vskip0.1cm
\noindent
iii) $k_s $ is constant with energy and $ k_{sh} $ starts decreasing with increasing energy, but asymptotically tends to a constant value, implying that  KNO scaling violation is not very strong. 

Our results suggest that none of these scenarios is supported by data. Moreover we emphasize the importance of  the relation (\ref{omegaf}) 
and of studying  the behavior of $ k_{total} $. Indeed,  since $\lambda$ and $\alpha$ may depend on  the energy, the simple constancy of  
$k_s $ and $ k_{sh} $ does not ensure KNO scaling. The constancy of $ k_{total} $  does. In Fig. \ref{fig:ktotal} we show $ k_{total} $ as a 
function of the energy for different pseudorapidity windows. It is constant for $|\eta| < 1.0$ and energy-dependent for $|\eta| > 1.0$.

We can compare our results with those found in 
\cite{ALICE:2015olq} where a similar DNBD fit was performed. The comparison shows that the order of magnitude of the parameters is the 
same. However, in contrast to \cite{ALICE:2015olq}, we find a clear trend which separates the narrower rapidity windows, 
$|\eta| < 1.0$,  from the wider ones, with $|\eta| > 1.0$. In the former  $k_s $ and $ k_{sh} $ are constant with energy, whereas in 
the latter these parameters decrease with increasing energies. This feature and the fact that in the narrower window we observe KNO scaling and not in the others strongly suggest that there is a change of dynamics when we go from one region to the other.  

To summarize:
\begin{enumerate}
\item Using the $k_T$ factorization formalism and introducing  a separation scale in the transverse momentum we were able to properly 
define soft and semihard multiplicities, giving a physical meaning to the previously called multiplicity components $n_1$ and $n_2$,  
proposed in \cite{Giovannini:1997ce} and used in \cite{ALICE:2015olq,acharya2017charged,acharya2023multiplicity}. 
\item With this separation, we could perform a DNBD fit of the multiplicy distributions measured at the LHC with fewer free parameters
(4 instead of 6),  obtaining very good fits. 
\item The resulting behavior of the parameters $\alpha$, $k_s$ and $k_{sh}$  with the collision energy is consistent (but slightly 
different) with those found in  previous analises. 
\item KNO scaling occurs for the total MD when $k_{total}$ is independent of the $\sqrt{s}$ and this happens only for $|\eta|<1.0$. Furthermore, within the same pseudorapidity window, KNO scaling is also satisfied for the soft and semihard components; that is, $k_s$ and $k_{sh}$ are also energy independent.

\item KNO violation is not related to changes in $\alpha$ and may have a dynamical origin due to changes in $N_s$, $N_{sh}$, $k_s$ and $k_{sh}$  with pseudorapidity.

\end{enumerate}

\eject 


\begin{acknowledgments}
We are grateful to M. Munhoz, G. Wilk and M.V.T. Machado for useful discussions. We were supported by the Brazilian funding agencies CAPES, FAPESP, CNPq and INCT-FNA. 
\end{acknowledgments}



\appendix

\section{Tables}

\begin{table}[h!]
\centering

\begin{minipage}{\textwidth}
\centering
\resizebox{\textwidth}{!}{
\begin{tabular}{|c|c|c|c|c|c|c|c|c|c|c|c|}
\hline
$\eta_c$ & data &$\sqrt{s}$ (TeV)& $\lambda$ & $\alpha$ & $\langle n \rangle_s$ & $\langle n \rangle_{sh}$ & $\langle n \rangle$ & $k_s$ & $ k_{sh}$ & $k_{total}$ & $\chi^2/\text{dof}$ \\
\hline

\multirow{5}{*}{0.5}
& \multirow{2}{*}{\cite{khachatryan2011charged}} &0.90 & 0.895 $ \pm $ 0.054 & 0.704 $ \pm $ 0.066 & 2.803 $ \pm $ 0.234 & 6.319 $ \pm $ 1.270 & 3.437  $ \pm $ 0.040 & 1.743 $ \pm $ 0.960 & 3.685 $ \pm $ 2.154 & 1.257 $ \pm $ 1.117 & 0.265 \\ 
&  &2.36 & 0.875 $ \pm $ 0.039 & 0.669 $ \pm $ 0.019 & 3.150 $ \pm $ 0.182 & 8.719 $ \pm $ 0.689 & 4.370 $ \pm $ 0.050 & 2.310 $ \pm $ 0.786 & 4.489 $ \pm $ 1.099 & 1.183 $ \pm $ 0.457 & 0.587 \\\cline{2-12}
& \cite{ALICE:2015olq} &2.76 & 0.936 $ \pm $ 0.011 & 0.661 $ \pm $ 0.010 & 3.028 $ \pm $ 0.065 & 8.599 $ \pm $ 0.273 & 4.599 $ \pm $ 0.057 & 2.061 $ \pm $ 0.336 & 3.770 $ \pm $ 0.279 & 1.249 $ \pm $ 0.168 & 0.057 \\\cline{2-12}
& \cite{khachatryan2011charged} &7.00 & 0.920 $ \pm $ 0.032 & 0.610 $ \pm $ 0.012 & 3.376 $ \pm $ 0.182 & 11.214 $ \pm $ 0.711 & 5.915 $ \pm $ 0.075 & 2.192 $ \pm $ 0.444 & 3.788 $ \pm $ 0.232 & 1.117 $ \pm $ 0.235 & 0.102 \\\cline{2-12}
& \cite{ALICE:2015olq}&8.00 & 0.955 $ \pm $ 0.009 & 0.594 $ \pm $ 0.008 & 3.270 $ \pm $ 0.055 & 10.733 $ \pm $ 0.219 & 6.018  $ \pm $ 0.049 & 1.882 $ \pm $ 0.263 & 3.344 $ \pm $ 0.152 & 1.164 $ \pm $ 0.108 & 0.106 \\
\hline

\multirow{2}{*}{0.8}
&\multirow{2}{*}{\cite{acharya2023multiplicity}}&5.02 & 1.072 $ \pm $ 0.006 & 0.640 $ \pm $ 0.002 & 4.195 $ \pm $ 0.041 & 13.634 $ \pm $ 0.130 & 8.135  $ \pm $ 0.052 & 1.890 $ \pm $ 0.054 & 3.652 $ \pm $ 0.054 & 1.519 $ \pm $ 0.073 & 0.615 \\
& &13.00 & 1.083 $ \pm $ 0.005 & 0.584 $ \pm $ 0.001 & 4.429 $ \pm $ 0.041 & 16.763 $ \pm $ 0.136 & 10.353  $ \pm $ 0.064 & 1.650 $ \pm $ 0.037 & 3.144 $ \pm $ 0.026 & 1.346 $ \pm $ 0.043 & 0.910 \\
\hline

\multirow{5}{*}{1.0}
& \multirow{2}{*}{\cite{khachatryan2011charged}}&0.90 & 0.891 $ \pm $ 0.021 & 0.730 $ \pm $ 0.009 & 5.514 $ \pm $ 0.176 & 14.160 $ \pm $ 0.680 & 6.987  $ \pm $ 0.081 & 2.552 $ \pm $ 0.388 & 6.096 $ \pm $ 0.745 & 1.405 $ \pm $ 0.289 & 0.254 \\
& &2.36 & 0.893 $ \pm $ 0.023 & 0.669 $ \pm $ 0.009 & 6.310 $ \pm $ 0.227 & 17.494 $ \pm $ 0.808 & 8.938 $ \pm $ 0.103 & 2.546 $ \pm $ 0.386 & 5.280 $ \pm $ 0.515 & 1.315 $ \pm $ 0.261 & 0.355 \\\cline{2-12}
& \cite{ALICE:2015olq}&2.76 & 0.936 $ \pm $ 0.009 & 0.662 $ \pm $ 0.007 & 6.176 $ \pm $ 0.122 & 17.594 $ \pm $ 0.442 & 9.393 $ \pm $ 0.116 & 2.268 $ \pm $ 0.178 & 4.195 $ \pm $ 0.190 & 1.315 $ \pm $ 0.111 & 0.040 \\\cline{2-12}
& \cite{khachatryan2011charged}&7.00 & 0.918 $ \pm $ 0.022 & 0.598 $ \pm $ 0.008 & 6.994 $ \pm $ 0.267 & 21.978 $ \pm $ 0.943 & 11.958 $ \pm $ 0.151 & 2.108 $ \pm $ 0.258 & 3.847 $ \pm $ 0.161 & 1.178 $ \pm $ 0.172 & 0.113 \\\cline{2-12}
& \cite{ALICE:2015olq}&8.00 & 0.947 $ \pm $ 0.008 & 0.600 $ \pm $ 0.005 & 6.608 $ \pm $ 0.104 & 22.196 $ \pm $ 0.392 & 12.157  $ \pm $ 0.098 & 2.151 $ \pm $ 0.129 & 3.767 $ \pm $ 0.109 & 1.187 $ \pm $ 0.074 & 0.080 \\
\hline

\multirow{5}{*}{1.5}
& \multirow{2}{*}{\cite{khachatryan2011charged}}&0.90 & 0.910 $ \pm $ 0.016 & 0.726 $ \pm $ 0.007 & 8.270 $ \pm $ 0.212 & 20.798 $ \pm $ 0.795 & 10.650  $ \pm $ 0.124 & 2.663 $ \pm $ 0.287 & 6.430 $ \pm $ 0.684 & 1.546 $ \pm $ 0.243 & 0.143 \\
& &2.36 & 0.925 $ \pm $ 0.020 & 0.662 $ \pm $ 0.009 & 9.427 $ \pm $ 0.299 & 25.370 $ \pm $ 1.044 & 13.704 $ \pm $ 0.158 & 2.325 $ \pm $ 0.295 & 5.010 $ \pm $ 0.578 & 1.429 $ \pm $ 0.264 & 0.180 \\\cline{2-12}
& \cite{ALICE:2015olq}&2.76 & 0.941 $ \pm $ 0.008 & 0.654 $ \pm $ 0.006 & 9.504 $ \pm $ 0.177 & 26.216 $ \pm $ 0.599 & 14.386 $ \pm $ 0.177 & 2.590 $ \pm $ 0.167 & 4.397 $ \pm $ 0.220 & 1.430 $ \pm $ 0.121 & 0.100 \\\cline{2-12}
& \cite{khachatryan2011charged}&7.00 & 0.918 $ \pm $ 0.018 & 0.591 $ \pm $ 0.007 & 10.723 $ \pm $ 0.347 & 32.745 $ \pm $ 1.153 & 18.097 $ \pm $ 0.228 & 2.168 $ \pm $ 0.211 & 3.994 $ \pm $ 0.139 & 1.229 $ \pm $ 0.151 & 0.188 \\\cline{2-12}
& \cite{ALICE:2015olq}&8.00 & 0.947 $ \pm $ 0.007 & 0.597 $ \pm $ 0.004 & 10.061 $ \pm $ 0.142 & 33.238 $ \pm $ 0.497 & 18.372  $ \pm $ 0.149 & 2.281 $ \pm $ 0.119 & 3.910 $ \pm $ 0.112 & 1.231 $ \pm $ 0.073 & 0.144 \\
\hline

\multirow{4}{*}{2.0}
& \multirow{3}{*}{\cite{khachatryan2011charged}}&0.90 & 0.909 $ \pm $ 0.015 & 0.719 $ \pm $ 0.006 & 11.233 $ \pm $ 0.269 & 27.433 $ \pm $ 0.939 & 14.344  $ \pm $ 0.167 & 2.895 $ \pm $ 0.300 & 6.679 $ \pm $ 0.696 & 1.633 $ \pm $ 0.251 & 0.167 \\
& &2.36 & 0.929 $ \pm $ 0.018 & 0.663 $ \pm $ 0.007 & 12.671 $ \pm $ 0.359 & 34.319 $ \pm $ 1.193 & 18.543 $ \pm $ 0.214 & 2.703 $ \pm $ 0.305 & 5.804 $ \pm $ 0.576 & 1.537 $ \pm $ 0.258 & 0.191 \\
& &7.00 & 0.908 $ \pm $ 0.016 & 0.583 $ \pm $ 0.006 & 14.731 $ \pm $ 0.447 & 43.493 $ \pm $ 1.400 & 24.252 $ \pm $ 0.306 & 2.257 $ \pm $ 0.211 & 4.089 $ \pm $ 0.146 & 1.256 $ \pm $ 0.147 & 0.310 \\\cline{2-12}
& \cite{acharya2017charged}&8.00 & 0.898 $ \pm $ 0.005 & 0.584 $ \pm $ 0.004 & 14.539 $ \pm $ 0.175 & 45.398 $ \pm $ 0.526 & 24.580  $ \pm $ 0.199 & 2.139 $ \pm $ 0.065 & 3.478 $ \pm $ 0.116 & 1.099 $ \pm $ 0.053 & 0.861 \\
\hline

\multirow{4}{*}{2.4}
& \multirow{3}{*}{\cite{khachatryan2011charged}}&0.90 & 0.913 $ \pm $ 0.013 & 0.718 $ \pm $ 0.005 & 13.464 $ \pm $ 0.296 & 32.726 $ \pm $ 1.001 & 17.245  $ \pm $ 0.200 & 3.069 $ \pm $ 0.270 & 7.321 $ \pm $ 0.677 & 1.716 $ \pm $ 0.228 & 0.180 \\
& &2.36 & 0.934 $ \pm $ 0.016 & 0.660 $ \pm $ 0.006 & 15.252 $ \pm $ 0.394 & 40.810 $ \pm $ 1.273 & 22.381 $ \pm $ 0.258 & 2.887 $ \pm $ 0.284 & 6.201 $ \pm $ 0.551 & 1.630 $ \pm $ 0.248 & 0.153 \\
& &7.00 & 0.907 $ \pm $ 0.015 & 0.578 $ \pm $ 0.006 & 17.846 $ \pm $ 0.504 & 51.731 $ \pm $ 1.520 & 29.133 $ \pm $ 0.367 & 2.303 $ \pm $ 0.202 & 4.132 $ \pm $ 0.154 & 1.284 $ \pm $ 0.141 & 0.411 \\\cline{2-12}
& \cite{acharya2017charged}&8.00 & 0.897 $ \pm $ 0.007 & 0.579 $ \pm $ 0.005 & 17.579 $ \pm $ 0.255 & 53.966 $ \pm $ 0.818 & 29.495  $ \pm $ 0.239 & 2.213 $ \pm $ 0.089 & 3.581 $ \pm $ 0.179 & 1.132 $ \pm $ 0.073 & 0.337 \\
\hline

\multirow{3}{*}{3.0}
& \multirow{3}{*}{\cite{acharya2017charged}}&0.90 & 0.860 $ \pm $ 0.006 & 0.692 $ \pm $ 0.006 & 19.265 $ \pm $ 0.303 & 41.306 $ \pm $ 0.993 & 22.409  $ \pm $ 0.253 & 3.247 $ \pm $ 0.147 & 4.750 $ \pm $ 0.428 & 1.503 $ \pm $ 0.126 & 2.322 \\
& &7.00 & 0.912 $ \pm $ 0.007 & 0.592 $ \pm $ 0.004 & 21.332 $ \pm $ 0.304 & 65.459 $ \pm $ 0.960 & 35.886 $ \pm $ 0.294 & 2.278 $ \pm $ 0.101 & 3.873 $ \pm $ 0.175 & 1.200 $ \pm $ 0.080 & 0.281 \\
& &8.00 & 0.896 $ \pm $ 0.007 & 0.574 $ \pm $ 0.005 & 22.050 $ \pm $ 0.313 & 66.437 $ \pm $ 0.954 & 36.684  $ \pm $ 0.297 & 2.291 $ \pm $ 0.093 & 3.711 $ \pm $ 0.195 & 1.170 $ \pm $ 0.077 & 0.311 \\
\hline

\multirow{3}{*}{3.4}
& \multirow{3}{*}{\cite{acharya2017charged}}&0.90 & 0.851 $ \pm $ 0.006 & 0.691 $ \pm $ 0.005 & 21.766 $ \pm $ 0.321 & 46.526 $ \pm $ 0.952 & 25.036  $ \pm $ 0.283 & 3.435 $ \pm $ 0.147 & 5.482 $ \pm $ 0.407 & 1.545 $ \pm $ 0.113 & 2.656 \\
& &7.00 & 0.916 $ \pm $ 0.007 & 0.587 $ \pm $ 0.004 & 24.058 $ \pm $ 0.346 & 72.604 $ \pm $ 1.059 & 40.405 $ \pm $ 0.331 & 2.310 $ \pm $ 0.102 & 3.998 $ \pm $ 0.189 & 1.250 $ \pm $ 0.082 & 0.216 \\
& &8.00 & 0.902 $ \pm $ 0.007 & 0.571 $ \pm $ 0.004 & 24.789 $ \pm $ 0.344 & 73.676 $ \pm $ 0.999 & 41.289  $ \pm $ 0.334 & 2.324 $ \pm $ 0.094 & 3.835 $ \pm $ 0.195 & 1.220 $ \pm $ 0.077 & 0.294 \\
\hline

\end{tabular}}
\end{minipage}\vspace{0.2cm}
\begin{minipage}{\textwidth}
\centering
\resizebox{\textwidth}{!}{
\begin{tabular}{|c|c|c|c|c|c|c|c|c|c|c|c|}
\hline
$\sqrt{s}$ (TeV) & data &$\eta_c$ & $\lambda$ & $\alpha$ & $\langle n \rangle_s$ & $\langle n \rangle_{sh}$ & $\langle n \rangle$ & $k_s$ & $k_{sh}$ & $k_{total}$ & $\chi^2/\text{dof}$ \\
\hline

\multirow{7}{*}{0.90}
& \multirow{5}{*}{\cite{khachatryan2011charged}} &0.5 & 0.895 $ \pm $ 0.054 & 0.704 $ \pm $ 0.066 & 2.803 $ \pm $ 0.234 & 6.319 $ \pm $ 1.270 & 3.437 $ \pm $ 0.040 & 1.743 $ \pm $ 0.960 & 3.685 $ \pm $ 2.154 & 1.257 $ \pm $ 1.117 & 0.265 \\
& &1.0 & 0.891 $ \pm $ 0.021 & 0.730 $ \pm $ 0.009 & 5.514 $ \pm $ 0.176 & 14.160 $ \pm $ 0.680 & 6.987 $ \pm $ 0.081 & 2.552 $ \pm $ 0.388 & 6.096 $ \pm $ 0.745 & 1.405 $ \pm $ 0.289 & 0.254 \\
& &1.5 & 0.910 $ \pm $ 0.016 & 0.726 $ \pm $ 0.007 & 8.270 $ \pm $ 0.212 & 20.798 $ \pm $ 0.795 & 10.650 $ \pm $ 0.124 & 2.663 $ \pm $ 0.287 & 6.430 $ \pm $ 0.684 & 1.546 $ \pm $ 0.243 & 0.143 \\
& &2.0 & 0.909 $ \pm $ 0.015 & 0.719 $ \pm $ 0.006 & 11.233 $ \pm $ 0.269 & 27.433 $ \pm $ 0.939 & 14.344 $ \pm $ 0.167 & 2.895 $ \pm $ 0.300 & 6.679 $ \pm $ 0.696 & 1.633 $ \pm $ 0.251 & 0.167 \\
& &2.4 & 0.913 $ \pm $ 0.013 & 0.718 $ \pm $ 0.005 & 13.464 $ \pm $ 0.296 & 32.726 $ \pm $ 1.001 & 17.245 $ \pm $ 0.200 & 3.069 $ \pm $ 0.270 & 7.321 $ \pm $ 0.677 & 1.716 $ \pm $ 0.228 & 0.180 \\ \cline{2-12}
&\multirow{2}{*}{\cite{acharya2017charged}} &3.0 & 0.860 $ \pm $ 0.006 & 0.692 $ \pm $ 0.006 & 19.265 $ \pm $ 0.303 & 41.306 $ \pm $ 0.993 & 22.409 $ \pm $ 0.253 & 3.247 $ \pm $ 0.147 & 4.750 $ \pm $ 0.428 & 1.503 $ \pm $ 0.126 & 2.322 \\
& &3.4 & 0.851 $ \pm $ 0.006 & 0.691 $ \pm $ 0.005 & 21.766 $ \pm $ 0.321 & 46.526 $ \pm $ 0.952 & 25.036 $ \pm $ 0.283 & 3.435 $ \pm $ 0.147 & 5.482 $ \pm $ 0.407 & 1.545 $ \pm $ 0.113 & 2.656 \\
\hline

\multirow{5}{*}{2.36}
& \multirow{5}{*}{\cite{khachatryan2011charged}}&0.5 & 0.875 $ \pm $ 0.039 & 0.669 $ \pm $ 0.019 & 3.150 $ \pm $ 0.182 & 8.719 $ \pm $ 0.689 & 4.370 $ \pm $ 0.050 & 2.310 $ \pm $ 0.786 & 4.489 $ \pm $ 1.099 & 1.183 $ \pm $ 0.457 & 0.587 \\
& &1.0 & 0.893 $ \pm $ 0.023 & 0.669 $ \pm $ 0.009 & 6.310 $ \pm $ 0.227 & 17.494 $ \pm $ 0.808 & 8.938 $ \pm $ 0.103 & 2.546 $ \pm $ 0.386 & 5.280 $ \pm $ 0.515 & 1.315 $ \pm $ 0.261 & 0.355 \\
& &1.5 & 0.925 $ \pm $ 0.020 & 0.662 $ \pm $ 0.009 & 9.427 $ \pm $ 0.299 & 25.370 $ \pm $ 1.044 & 13.704 $ \pm $ 0.158 & 2.325 $ \pm $ 0.295 & 5.010 $ \pm $ 0.578 & 1.429 $ \pm $ 0.264 & 0.180 \\
& &2.0 & 0.929 $ \pm $ 0.018 & 0.663 $ \pm $ 0.007 & 12.671 $ \pm $ 0.359 & 34.319 $ \pm $ 1.193 & 18.543 $ \pm $ 0.214 & 2.703 $ \pm $ 0.305 & 5.804 $ \pm $ 0.576 & 1.537 $ \pm $ 0.258 & 0.191 \\
& &2.4 & 0.934 $ \pm $ 0.016 & 0.660 $ \pm $ 0.006 & 15.252 $ \pm $ 0.394 & 40.810 $ \pm $ 1.273 & 22.381 $ \pm $ 0.258 & 2.887 $ \pm $ 0.284 & 6.201 $ \pm $ 0.551 & 1.630 $ \pm $ 0.248 & 0.153 \\
\hline

\multirow{3}{*}{2.76}
& \multirow{3}{*}{\cite{ALICE:2015olq}}&0.5 & 0.936 $ \pm $ 0.011 & 0.661 $ \pm $ 0.010 & 3.028 $ \pm $ 0.065 & 8.599 $ \pm $ 0.273 & 4.599 $ \pm $ 0.057 & 2.061 $ \pm $ 0.336 & 3.770 $ \pm $ 0.279 & 1.249 $ \pm $ 0.168 & 0.057 \\
& &1.0 & 0.936 $ \pm $ 0.009 & 0.662 $ \pm $ 0.007 & 6.176 $ \pm $ 0.122 & 17.594 $ \pm $ 0.442 & 9.393 $ \pm $ 0.116 & 2.268 $ \pm $ 0.178 & 4.195 $ \pm $ 0.190 & 1.315 $ \pm $ 0.111 & 0.040 \\
& &1.5 & 0.941 $ \pm $ 0.008 & 0.654 $ \pm $ 0.006 & 9.504 $ \pm $ 0.177 & 26.216 $ \pm $ 0.599 & 14.386 $ \pm $ 0.177 & 2.590 $ \pm $ 0.167 & 4.397 $ \pm $ 0.220 & 1.430 $ \pm $ 0.121 & 0.100 \\
\hline

\multirow{1}{*}{5.02}
& \cite{acharya2023multiplicity}&0.8 & 1.072 $ \pm $ 0.006 & 0.640 $ \pm $ 0.002 & 4.195 $ \pm $ 0.041 & 13.634 $ \pm $ 0.130 & 8.135 $ \pm $ 0.052 & 1.890 $ \pm $ 0.054 & 3.652 $ \pm $ 0.054 & 1.519 $ \pm $ 0.073 & 0.615 \\
\hline

\multirow{7}{*}{7.00}
& \multirow{5}{*}{\cite{khachatryan2011charged}}&0.5 & 0.920 $ \pm $ 0.032 & 0.610 $ \pm $ 0.012 & 3.376 $ \pm $ 0.182 & 11.214 $ \pm $ 0.711 & 5.915 $ \pm $ 0.075 & 2.192 $ \pm $ 0.444 & 3.788 $ \pm $ 0.232 & 1.117 $ \pm $ 0.235 & 0.102 \\
& &1.0 & 0.918 $ \pm $ 0.022 & 0.598 $ \pm $ 0.008 & 6.994 $ \pm $ 0.267 & 21.978 $ \pm $ 0.943 & 11.958 $ \pm $ 0.151 & 2.108 $ \pm $ 0.258 & 3.847 $ \pm $ 0.161 & 1.178 $ \pm $ 0.172 & 0.113 \\
& &1.5 & 0.918 $ \pm $ 0.018 & 0.591 $ \pm $ 0.007 & 10.723 $ \pm $ 0.347 & 32.745 $ \pm $ 1.153 & 18.097 $ \pm $ 0.228 & 2.168 $ \pm $ 0.211 & 3.994 $ \pm $ 0.139 & 1.229 $ \pm $ 0.151 & 0.188 \\
& &2.0 & 0.908 $ \pm $ 0.016 & 0.583 $ \pm $ 0.006 & 14.731 $ \pm $ 0.447 & 43.493 $ \pm $ 1.400 & 24.252 $ \pm $ 0.306 & 2.257 $ \pm $ 0.211 & 4.089 $ \pm $ 0.146 & 1.256 $ \pm $ 0.147 & 0.310 \\
& &2.4 & 0.907 $ \pm $ 0.015 & 0.578 $ \pm $ 0.006 & 17.846 $ \pm $ 0.504 & 51.731 $ \pm $ 1.520 & 29.133 $ \pm $ 0.367 & 2.303 $ \pm $ 0.202 & 4.132 $ \pm $ 0.154 & 1.284 $ \pm $ 0.141 & 0.411 \\ \cline{2-12}
& \multirow{2}{*}{\cite{acharya2017charged}}&3.0 & 0.912 $ \pm $ 0.007 & 0.592 $ \pm $ 0.004 & 21.332 $ \pm $ 0.304 & 65.459 $ \pm $ 0.960 & 35.886 $ \pm $ 0.294 & 2.278 $ \pm $ 0.101 & 3.873 $ \pm $ 0.175 & 1.200 $ \pm $ 0.080 & 0.281 \\
& &3.4 & 0.916 $ \pm $ 0.007 & 0.587 $ \pm $ 0.004 & 24.058 $ \pm $ 0.346 & 72.604 $ \pm $ 1.059 & 40.405 $ \pm $ 0.331 & 2.310 $ \pm $ 0.102 & 3.998 $ \pm $ 0.189 & 1.250 $ \pm $ 0.082 & 0.216 \\
\hline

\multirow{7}{*}{8.00}
& \multirow{3}{*}{\cite{ALICE:2015olq}}&0.5 & 0.955 $ \pm $ 0.009 & 0.594 $ \pm $ 0.008 & 3.270 $ \pm $ 0.055 & 10.733 $ \pm $ 0.219 & 6.018 $ \pm $ 0.049 & 1.882 $ \pm $ 0.263 & 3.344 $ \pm $ 0.152 & 1.164 $ \pm $ 0.108 & 0.106 \\
& &1.0 & 0.947 $ \pm $ 0.008 & 0.600 $ \pm $ 0.005 & 6.608 $ \pm $ 0.104 & 22.196 $ \pm $ 0.392 & 12.157 $ \pm $ 0.098 & 2.151 $ \pm $ 0.129 & 3.767 $ \pm $ 0.109 & 1.187 $ \pm $ 0.074 & 0.080 \\
& &1.5 & 0.947 $ \pm $ 0.007 & 0.597 $ \pm $ 0.004 & 10.061 $ \pm $ 0.142 & 33.238 $ \pm $ 0.497 & 18.372 $ \pm $ 0.149 & 2.281 $ \pm $ 0.119 & 3.910 $ \pm $ 0.112 & 1.231 $ \pm $ 0.073 & 0.144 \\ \cline{2-12}
& \multirow{4}{*}{\cite{acharya2017charged}}&2.0 & 0.898 $ \pm $ 0.005 & 0.584 $ \pm $ 0.004 & 14.539 $ \pm $ 0.175 & 45.398 $ \pm $ 0.526 & 24.580 $ \pm $ 0.199 & 2.139 $ \pm $ 0.065 & 3.478 $ \pm $ 0.116 & 1.099 $ \pm $ 0.053 & 0.861 \\
& &2.4 & 0.897 $ \pm $ 0.007 & 0.579 $ \pm $ 0.005 & 17.579 $ \pm $ 0.255 & 53.966 $ \pm $ 0.818 & 29.495 $ \pm $ 0.239 & 2.213 $ \pm $ 0.089 & 3.581 $ \pm $ 0.179 & 1.132 $ \pm $ 0.073 & 0.337 \\
& &3.0 & 0.896 $ \pm $ 0.007 & 0.574 $ \pm $ 0.005 & 22.050 $ \pm $ 0.313 & 66.437 $ \pm $ 0.954 & 36.684 $ \pm $ 0.297 & 2.291 $ \pm $ 0.093 & 3.711 $ \pm $ 0.195 & 1.170 $ \pm $ 0.077 & 0.311 \\
& &3.4 & 0.902 $ \pm $ 0.007 & 0.571 $ \pm $ 0.004 & 24.789 $ \pm $ 0.344 & 73.676 $ \pm $ 0.999 & 41.289 $ \pm $ 0.334 & 2.324 $ \pm $ 0.094 & 3.835 $ \pm $ 0.195 & 1.220 $ \pm $ 0.077 & 0.294 \\
\hline

\multirow{1}{*}{13.00}
& \cite{acharya2023multiplicity}&0.8 & 1.083 $ \pm $ 0.005 & 0.584 $ \pm $ 0.001 & 4.429 $ \pm $ 0.041 & 16.763 $ \pm $ 0.136 & 10.353 $ \pm $ 0.064 & 1.650 $ \pm $ 0.037 & 3.144 $ \pm $ 0.026 & 1.346 $ \pm $ 0.043 & 0.910 \\
\hline

\end{tabular}}
\end{minipage}
\caption{Parameters  obtained from the  fit of data with the double NBD (\ref{DNBD}) for different pseudorapidity  windows 
($- \eta_c < \eta < + \eta_c$) and energies ($\sqrt{s}$). The data used for the fits are from CMS \cite{khachatryan2011charged} and ALICE \cite{ALICE:2015olq,acharya2017charged,acharya2023multiplicity}. The break in trend in certain parameters, such as $k_{sh}$, for the
energies $0.9$, $7.0$  and $8$ TeV  arises primarily because the data used for the fitting originate from different studies, which may have 
employed different  methodologies or approaches.}
\label{tab-NBD}
\end{table}

\end{document}